\documentclass[manuscript]{acmart}
\usepackage{color}
\usepackage{multirow}
\usepackage{makecell}
\usepackage{subfig}
\usepackage{colortbl}
\usepackage{xcolor}
\usepackage{pgfplots}
\usepackage{pgfplotstable}
\usepackage{appendix}
\usepackage{graphicx}
\usepackage{array}
\pgfplotsset{compat=1.18}
\pgfplotsset{compat=1.18}
\usepackage{caption}
\usepackage{enumitem}
\usepackage{longtable}
\usepackage{multirow}
\usepackage{booktabs}

\author{Yibo Meng}
\affiliation{%
  \institution{Tsinghua University}
  \city{Beijing}
  \country{China}
}
\email{mengyb22@mails.tsinghua.edu.cn}

\author{Rong Fu}
\affiliation{%
  \institution{The Institute of Collaborative Innovation}
  \city{Macau}
  \country{China}
}
\email{mc46603@um.edu.mo}

\author{Lyumanshan Ye}
\affiliation{%
  \institution{Shanghai Jiao Tong University}
  \city{Shanghai}
  \country{China}
}
\email{yelyumanshan@sjtu.edu.cn}

\author{Zhiming Liu}
\affiliation{%
  \institution{University of Shanghai for Science and Technology}
  \city{Shanghai}
  \country{China}
}

\author{Zhixin Cai}
\affiliation{%
  \institution{Zhejiang University of Science and Technology}
  \city{Hangzhou}
  \country{China}
}
\email{xinlise@gmail.com}

\author{Xiaolan Ding}
\affiliation{%
  \institution{North China University of Science and Technology Health Science Center}
  \city{Tangshan}
  \country{China}
}

\author{Yan Guan}
\affiliation{%
  \institution{Arts \& Design Academy, Tsinghua University}
  \city{Beijing}
  \country{China}
}
\email{guany@tsinghua.edu.cn}

\begin{document}

\title{Final Happiness: What Intelligent User Interfaces Can Do for The Lonely Dying}

\renewcommand{\shortauthors}{Meng et al.}

\begin{abstract}
This study explores the design of Intelligent User Interfaces (IUIs) to address the profound existential loneliness of terminally ill individuals. While Human-Computer Interaction (HCI) has made inroads in "Thanatechnology," current research often focuses on practical aspects like digital legacy management, overlooking the subjective, existential needs of those facing death in isolation. To address this gap, we conducted in-depth qualitative interviews with 14 lonely, terminally ill individuals. Our core contributions are: (1) An empirically-grounded model articulating the complex psychological, practical, social, and spiritual needs of this group; (2) The "Three Pillars, Twelve Principles" framework for designing IUIs as "Existential Companions"; and (3) A critical design directive derived from user evaluations: technology in this context should aim for transcendence over simulation. The findings suggest that IUIs should create experiences that augment or surpass human capabilities, rather than attempting to simulate basic human connections, which can paradoxically deepen loneliness. This research provides a clear, user-centered path for designing technology that serves not as a "tool for dying," but as a "partner for living fully until the end".
\end{abstract}

\maketitle

\section{Introduction}

The end of life is not merely a biological event but a profound psychosocial and existential experience~\cite{massimi2011matters, albers2023dying}. While modern medicine has made significant strides in extending human lifespan, it has not equally equipped individuals to find meaning and connection in their final chapter~\cite{Byock2025}. When diagnosed with a terminal illness, individuals confront not only physical decline and the fear of mortality but also a more insidious "second crisis": a profound sense of loneliness~\cite{ferguson2014craving}.

This is not simple social isolation but a deeper, existential loneliness—a feeling of disconnection from oneself, from others, and from a sense of meaning in the world. Empirical research underscores the severity of this issue. In a survey of palliative care professionals, 92.6\% perceived loneliness as highly prevalent among terminally ill patients. This loneliness is often a "taboo subject," difficult to articulate yet devastating to a patient's well-being~\cite{abernethy2010strategy}, with studies indicating a significant correlation between loneliness and the exacerbation of physical symptoms like pain and breathlessness, as well as severe detriment to psychological, spiritual, and social welfare~\cite{sandham2022intelligent}. Addressing end-of-life loneliness has thus become an urgent and formidable challenge in modern palliative care~\cite{ritchie2017better, plotzke2021construction}.

In recent years, an emerging subfield within Human-Computer Interaction (HCI)—"Thanatechnology"~\cite{MassimiCharise2009}—has begun to explore how technology can intervene in the contexts of death, dying, and bereavement~\cite{albers2023dying, massimi2011matters, Massimi2010}. Early work in this area has yielded valuable results, primarily focusing on resolving the practical problems associated with mortality. For instance, a significant body of research has investigated the management and inheritance of digital legacies, such as social media accounts, digital photos, and personal data~\cite{chen2021happens, odom2010passing}. Other work has focused on providing support for grieving families and social networks~\cite{massimi2011dealing, massimi2013exploring, kim2024maintaining, MassimiBaecker2010} or developing coordination tools for informal caregivers~\cite{saito2025unintended}.

While invaluable, these studies have largely concentrated on the "practicalities" of death (e.g., asset management, post-mortem memorialization) or served the social ecosystem surrounding the patient (e.g., family, caregivers). However, the HCI community itself has recognized the limitations of this focus. As articulated by the "HCI at End of Life \& Beyond" workshop, there is a pressing need for researchers to "go far further to embrace the contexts relating to [death] more meaningfully and broadly"~\cite{wallace2020hci}. This signals a disciplinary shift from a concern with the \textit{affairs of the dead} to a deeper engagement with the \textit{experience of the dying}~\cite{ahmadpour2023can}.

The central thesis of this paper is that a significant gap exists in current HCI research: we lack designs that directly address the \textit{subjective, existential experience} of individuals who are isolated from traditional human support networks and are facing the end of life in solitude. Existing technologies may help them manage their affairs or connect with others~\cite{tan2024scoping, Stanley2024}, but few are designed to accompany them in confronting existential voids, searching for ultimate meaning, or affirming their self-worth amidst the fear of being forgotten~\cite{smith2024designing}.

This gap stems not from a lack of technical capability but from a misdiagnosis of the problem's nature. The core of end-of-life loneliness is often not the physical absence of social contact but a profound crisis of meaning. The primary needs expressed by our study's participants—to combat "existential void," seek "ultimate confirmation of meaning," desire to be "witnessed," and explore "spiritual transcendence"—are not functional tasks but deep calls for existential companionship. Therefore, merely providing connectivity tools like video chat or social media is insufficient~\cite{Ostherr2016}, as these tools incorrectly treat an existential problem as a social one.

Based on this, this study poses the following core research question: \textit{How should Intelligent User Interfaces (IUIs) be designed to transcend the role of functional aids and become companions capable of accompanying individuals through complex psychological, emotional, social, existential, and spiritual domains, especially as they face death in solitude?} To answer this question, this study employed a rigorous and ethical qualitative research methodology, conducting in-depth, semi-structured interviews with 14 terminally ill individuals experiencing profound loneliness. Our research design prioritized participant welfare over knowledge acquisition, implementing exhaustive ethical safeguards~\cite{Abejas2025} to ensure that our engagement with this extremely vulnerable group was safe, respectful, and meaningful.

The core contributions of this study are threefold:

\begin{enumerate}
    \item \textbf{An Empirically-Grounded Model of Existential Needs:} This study is the first to systematically articulate the complex needs of lonely, terminally ill individuals across four dimensions: psychological-emotional, practical-control, social-existential, and spiritual-transcendental.
    \item \textbf{An Innovative Framework for Designing "Existential Companions":} Based on this needs model, we propose a design framework titled "Three Pillars, Twelve Principles." This framework provides concrete guidance for designing IUIs that can act as both "proactive life custodians" and "empathetic companions." Its core principles, such as "Presumed Empathy," "Life Narrative and Eternal Witness," and "Sacred Data and Legacy Stewardship," are intended to elevate technical ethics, system functionality, and humanistic care to equal importance.
    \item \textbf{A Critical Design Directive for Thanatechnology:} Through user evaluation of our prototype concepts, this study derives a crucial design insight. Our findings indicate that in the context of end-of-life care, the value of technology lies not in \textbf{simulating} the basic emotional connections that humans are meant to provide (a simulation that paradoxically highlights their absence), but in \textbf{creating transcendent} experiences—that is, augmenting or surpassing what an individual can achieve given their physical or social limitations. This finding offers a clear, user-centered, and critical directive for the entire field of Thanatechnology, pushing it to evolve from designing "tools for dying" toward designing "partners for living fully until the end."
\end{enumerate}

\section{Related Work}

To clearly situate the unique contribution of this study, we review and analyze existing literature from four key areas of HCI research: methodological approaches in end-of-life HCI, technologies designed for the end-of-life social ecosystem, digital and physical legacy management, and technological interventions aimed at alleviating loneliness.

\subsection{The Landscape of HCI at End of Life: Methodological and Ethical Frontiers}

One of the core challenges for "Thanatechnology" as a research field is how to ethically and productively engage with extremely vulnerable populations~\cite{ahmadpour2023can}. Foundational work has called for the design of "thanatosensitive" technologies~\cite{MassimiCharise2009} to support practices like remembrance and the bequeathing of digital assets~\cite{massimi2013exploring, odom2010passing, MassimiBaecker2011}.

The academic community widely acknowledges that this research area is "fraught with ethical and methodological dilemmas"~\cite{ahmadpour2023can}. Designing for vulnerable people, such as those in advanced dementia care~\cite{beijer2025mano}, requires researchers to be highly sensitive to the risks of potential stigmatization and aware of the challenges inherent in deep participant involvement~\cite{Abejas2025}. This demands not only robust professional skills but also profound humanistic care. Therefore, the exhaustive ethical review, specialized researcher training, and multiple participant protection mechanisms employed in this study are not merely prerequisites; they are a response to the field's methodological inquiries, demonstrating how deep and authentic user insights can be obtained while ensuring participant dignity and safety. Emerging design approaches focusing on user-centered and iterative participatory design further signal the field's move towards more proactive and sensitive design methodologies~\cite{abdel2022development, blanes2023user, ullal2024iterative, carey2023co}.

\subsection{Designing for the End-of-Life Social Ecosystem}

A significant branch of existing research focuses on the social support network surrounding the patient—the ecosystem of family, friends, informal caregivers, and healthcare professionals~\cite{ferguson2014craving, suslow2023impact, wu2024collective}. For example, the work of Saito and Sugihara introduces the concept of "Unintended, Percolated Work" (UPW), which powerfully reveals the dilemmas arising from misaligned expectations and poor communication between family caregivers and medical professionals~\cite{saito2025unintended}. Their research identifies how this misalignment leads to "overloaded work" and "overlooked work" for caregivers, as well as "overstepped work" by medical staff, ultimately causing immense psychological distress and regret for all parties involved~\cite{saito2025unintended}. To mitigate such issues, researchers have explored various digital health interventions and interdisciplinary information tools to improve collaboration and information flow in palliative care contexts~\cite{kuziemsky2008interdisciplinary, jeon2025targeted, kuziemsky2006hospice, cox2018palliative}.

This body of literature provides a crucial social context for our study. The work of Saito and Sugihara helps explain \textit{why} terminally ill individuals may fall into loneliness—because the very social care system meant to support them is itself fragile and fraught with tension~\cite{saito2025unintended}. However, the focus of such research is on improving collaboration \textit{among} caregivers and medical professionals. Predictive modeling and informatics are also being leveraged to identify patients who may benefit from palliative care, but this again focuses on the clinical delivery system~\cite{murphree2021improving, courtright2019electronic}. Our study, in contrast, turns its attention to those individuals who are already at the margins of this ecosystem, or have detached from it entirely. For them, the problem is not one of improving collaboration but of navigating the journey alone in an environment with little to no human support. Thus, the IUI proposed in our research is not intended to replace human care~\cite{Vandersman2024}, but to provide a necessary and critical form of supplemental support when the human social ecosystem is absent or has failed.

\subsection{Digital and Physical Legacy: Crafting a Posthumous Identity}

Digital legacy management is one of the most developed areas within Thanatechnology~\cite{albers2023dying}. Systematic literature reviews have outlined the field's core themes, including the continuity of post-mortem identity, audience and access rights for digital assets, and the care and disposition of digital remains~\cite{chen2021happens, odom2010passing}. However, significant divergence exists on how to best approach digital legacy planning.

Some research has critiqued prevalent "asset-first, inventory-centric" planning methods, finding that asking users to list and itemize their digital assets often leaves them feeling overwhelmed. In contrast, many advocate for a more facilitated, mission-driven, and relationship-centric approach, which begins with a user's core values and significant relationships to determine what digital content is worth preserving~\cite{chen2021happens, kim2024maintaining}.

This finding resonates strongly with the insights from our study. Our participants universally expressed a powerful desire to become "storytellers," seeking "ultimate confirmation of meaning" by repeatedly narrating and reinterpreting their lives. This is a quintessential values-centric, rather than asset-centric, act of legacy construction. Accordingly, the IUI prototype concepts we propose, such as the "Life Memoir Creator" and "Meaning of Life Explorer," can be seen as technological implementations of a "relationship-centric" philosophy. They transform legacy planning from a daunting post-mortem task into a deeply therapeutic process of life review conducted in the present~\cite{massimi2011dealing}. This reflects a core tenet of our research: to seamlessly merge the design for \textit{memory} (digital legacy) with the design for the \textit{present moment} (the end-of-life experience), enriching and settling the final moments of life by reflecting on the past and planning for the future.

\subsection{Positioning This Study's Contribution}

In summary, this study is uniquely positioned at the intersection of several critical research gaps. As illustrated in Table 1, previous research has separately addressed the \textbf{social ecosystem} of the dying~\cite{saito2025unintended}, their \textbf{post-mortem digital artifacts}~\cite{chen2021happens, odom2010passing}, and \textbf{solutions for social loneliness} using various technologies~\cite{maguraushe2024use, tan2024scoping}. While AI and IUIs are being explored for clinical decision support and art therapy~\cite{narvaez2025artificial, kim2023alphadapr}, few studies have translated a deep understanding of the terminally ill individual's inner \textit{existential needs} directly into a robust, ethical, and user-validated design framework for an intelligent system.

This study fills this gap by: first, constructing a multi-dimensional model of existential needs through rigorous qualitative research; second, proposing a set of principles and concepts aimed at designing an "existential companion" rather than a "functional tool"~\cite{narvaez2025artificial}; and finally, distilling a core design directive of "transcendence over simulation" through user evaluation. The contribution of this paper is to provide a complete pathway for the HCI field—from deep empathy to concrete design, and from theory-building to practical validation—in addressing the ultimate challenge of end-of-life loneliness.

\begin{table*}[ht]
\centering
\caption{Positioning This Study within the HCI Research Landscape at End of Life}
\label{tab:positioning}
\resizebox{\textwidth}{!}{%
\begin{tabular}{|p{0.2\textwidth}|p{0.2\textwidth}|p{0.2\textwidth}|p{0.2\textwidth}|p{0.2\textwidth}|}
\hline
\textbf{Research Stream} & \textbf{Primary Focus} & \textbf{Core Concepts} & \textbf{Limitation / Unaddressed Gap} & \textbf{This Study's Contribution} \\
\hline
\textbf{Social Ecosystem Support} (e.g., Saito \& Sugihara~\cite{saito2025unintended}, Kuziemsky et al.~\cite{kuziemsky2008interdisciplinary}) & Caregiver-professional-family network & Collaboration, communication gaps, "unintended work", clinical informatics & Primarily focuses on the support network, not the internal, existential experience of the isolated individual. & Provides a necessary support solution for individuals when the social ecosystem is absent or has failed. \\
\hline
\textbf{Digital \& Physical Legacy} (e.g., Chen et al.~\cite{chen2021happens}, Kim et al.~\cite{kim2024maintaining}) & Post-mortem digital assets and physical artifacts & Values-based planning, identity, memory, artifact lifecycle & Often treats legacy as a future-oriented task, not as a present-tense, therapeutic process. & Reframes legacy creation as a present-moment tool for meaning-making and life review at the end of life. \\
\hline
\textbf{Loneliness \& Social Connection Tech} (e.g., gerontechnology) & Mitigating social isolation through connection and companionship & Social robots, VR~\cite{deng2025research}, smart speakers~\cite{doganguen2025speech}, simulating presence & Hollow simulation risks deepening loneliness; often misdiagnoses existential loneliness as social isolation. & Provides an evidence-based directive to shift design from "simulation" to "transcendence," and from social partner to existential companion. \\
\hline
\textbf{This Study} & \textbf{The direct, subjective, existential experience of the lonely, terminally ill individual} & \textbf{Existential companion, meaning-making, transcendence, micro-control, sacred data} & & Integrates the above themes into a holistic, empirically-grounded framework for designing an "existential companion." \\
\hline
\end{tabular}%
}
\end{table*}
\section{Preliminary Qualitative Interviews}
This study employed a qualitative research design. Through in-depth, semi-structured interviews, we explored the psychological needs of individuals facing end-of-life loneliness and their acceptance of and expectations for Intelligent User Interfaces (IUIs).

\subsection{Participants}
We visited 12 public tertiary hospitals to seek support and recruit participants, ultimately receiving support from Anonymous Hospital.

\textbf{Inclusion Criteria} (must meet all): Diagnosed with a terminal illness (e.g., end-stage organ failure) by a physician with a limited life expectancy; in a long-term state of loneliness (no surviving immediate family or no visits from immediate family/close friends); physician's initial assessment indicates physical and cognitive suitability for the study, followed by a clinical cognitive assessment by the researchers; full understanding of the research purpose, procedures, and potential impacts, and signing a voluntary consent form.

\textbf{Exclusion Criteria} (meet any): Cognitive impairment or mental illness affecting understanding and judgment as determined by a physician; physical condition deemed unsuitable for the study by a physician; severe hearing or speech impairments; currently in emotional distress, agitation, or psychological crisis where participation could cause harm, as assessed by a physician; deep-seated resistance, fear, or aversion to smart devices or AI, preventing objective discussion of IUI-related topics.

\textbf{Ethics}: This study received dual ethical approval from Anonymous University and Anonymous Hospital. We acknowledge that the participants are an extremely vulnerable group. Therefore, the study design prioritized participant welfare over knowledge acquisition, strictly adhering to the ethical principles of respect, beneficence, justice, and non-maleficence. Specific considerations included:
\begin{enumerate} 
    \item \textbf{Participant Screening}: Initially, only individuals confirmed by a physician to have unimpaired consciousness and no severe physical pain under current medical intervention were considered. Before any contact, the attending physician performed a secondary screening for cognitive impairment and physical stability. Only those deemed "stable and suitable for participation" were included. For those who passed, researchers used an adapted version of the MacArthur Competence Assessment Tool principles to assess, through conversation, their ability to: 1) understand the study's purpose, procedures, and risks; 2) relate this information to their own situation; 3) reason logically; and 4) express a clear choice. This was recorded. Only those with full comprehension proceeded. We used clear, simple, jargon-free language in the consent form and a "step-by-step explanation" approach, having participants paraphrase key points to ensure understanding. Researchers reiterated the purpose, procedures, potential impacts, and participant rights at least three times before the interview. Participants' rights included: strict confidentiality and anonymity; the right to review or delete their data; the right to skip questions or pause at any time; the right to withdraw without reason; the right to ask researchers to leave; the right to have the study purpose restated; and immediate access to their doctors, nurses, and a collaborating psychologist.
    \item \textbf{Minimizing Psychological Risks and Providing Immediate Support}: Acknowledging that discussing end-of-life experiences can cause distress, all interviewers received specialized training in palliative care communication, grief counseling, active listening, and crisis recognition (see Appendix A). A senior psychological counselor was on standby during interviews. If significant distress occurred, the interviewer would pause and offer professional support. A mental health support hotline was also provided for delayed emotional reactions.
    \item \textbf{Ensuring Privacy and Confidentiality}: All recordings were immediately uploaded to an encrypted, password-protected server and deleted after transcription. Transcripts were anonymized, with all personal identifiers removed. In publications, non-critical details that could identify participants were merged or obscured. Only core research team members had access to non-anonymized data.
    \item \textbf{Balancing Benefits and Burdens}: We were transparent that direct medical benefits were limited. Potential benefits included an opportunity for a listened-to life review and meaning-making. The study's social value lies in providing empirical evidence to improve psychosocial support for this group. Each participant received a bouquet of flowers as a token of appreciation. Each participating physician and psychologist was compensated with 100 RMB.
\end{enumerate}

\subsection{Procedure}
We conducted face-to-face, semi-structured interviews following the procedure detailed in Table \ref{tab:interview_procedure}.

\begin{table*} h!
\centering
    \caption{Detailed Interview Procedure}
\label{tab:interview_procedure}
\begin{tabular}{|p{0.15\linewidth}|p{0.1\linewidth}|p{0.2\linewidth}|p{0.45\linewidth}|}
    \hline
    \textbf{Phase} & \textbf{Task Code} & \textbf{Task Name} & \textbf{Core Operations and Checkpoints} \\
    \hline
    \multirow{3}{=}{Pre-Interview Preparation} & T1 & Participant Selection & Confirm the day's participant from the pre-approved and consented list. Re-confirm the participant's current physical and mental suitability with their doctor before the interview. \\
    \cline{2-4}
    & T2 & Environment Preparation & Ensure the interview room is private, well-ventilated, with soft lighting. Ask about temperature, lighting, or background music preferences. Ensure the emergency call bell is accessible. Check all recording equipment. Resolve any environmental issues or cancel if necessary. \\
    \cline{2-4}
    & T3 & Interviewer Self-Preparation & Conduct a brief meditation. Review participant's basic info and the interview guide. Re-read interviewer guidelines (Appendix A). Notify the on-standby doctor, nurse, and psychologist that the interview is starting. \\
    \hline
    \multirow{3}{=}{Interview Execution} & T4 & Initial Contact \& Rapport Building & Enter quietly after knocking. Make a brief introduction and engage in light conversation. Reiterate the study purpose, procedure, potential impacts, and all participant rights. Start recording only after explicit consent. \\
    \cline{2-4}
    & T5 & Core Interview Phase & Proceed through the six modules of the semi-structured questionnaire (Appendix B). Use open-ended questions. Provide non-verbal support. Use validating language for expressed pain. Allow for valuable digressions. Actively monitor for signs of fatigue and offer breaks. \\
    \cline{2-4}
    & T6 & Conclusion \& Emotional Transition & Signal the end is approaching. Express sincere gratitude using the scripted text, emphasizing the value of their story. Turn off recording equipment. Re-check the participant's state, ensure call bell and water are at hand, tidy the room, and leave quietly. \\
    \hline
    \multirow{3}{=}{Post-Interview Processing} & T7 & Debriefing & Immediately upload the recording to the secure server. Briefly report the participant's emotional state to the standby psychologist and research lead. Conduct a personal emotional debrief, seeking support if needed to process potential vicarious trauma. \\
    \cline{2-4}
    & T8 & Transcription \& Anonymization & Use a professional transcriber under a confidentiality agreement. Review the transcript to remove all identifiers and obscure potentially identifying details. Securely delete the original audio file after verification. \\
    \cline{2-4}
    & T9 & Long-term Participant Welfare & Participants retain the right to request data deletion even after the interview. Maintain open lines of communication. \\
    \hline
\end{tabular}
\end{table*}

\subsection{Interview Questionnaire Design}
The interview questionnaire (see Appendix B) was designed with a progressive narrative logic: from distant to near, from surface to core, from past to future, and from expression to empowerment, guiding participants through a safe, deep, and meaningful life narrative journey.
The opening section establishes a safe space by reiterating rights and empowering the participant, building trust for the subsequent in-depth dialogue. The questionnaire then begins with (1) \textbf{Building Empathetic Connection}, inviting participants to recall cherished memories and roles, treating them as whole individuals with valuable stories, not just as "patients".
Following this, the question sequence naturally flows to (2) \textbf{Exploring the Inner World}. It starts by assessing their current emotional state, then explores specific fears, and subsequently uncovers their inner resources and positive moments to avoid dwelling in despair. This is followed by a deeper dive into their definition of loneliness and their deepest needs for connection, before addressing unfinished business, regrets, and dignity.
The questionnaire then creates a closed loop from psychological needs to practical care. After exploring the inner world, (3) \textbf{Seeking Solace} and (4) \textbf{Specific Wishes} modules translate abstract emotional needs into concrete sources of comfort and achievable wishes (e.g., tasting a specific food again).
The design then skillfully integrates traditional humanistic care with forward-looking technology. (5) \textbf{Views on Intelligent User Interfaces (IUI)} module is introduced only after a comprehensive understanding of the participant's loneliness, connection needs, and wishes has been established, guiding them to consider how technology might address these revealed needs.
Finally, the questionnaire concludes with empowerment. The (6) \textbf{Messages and Feedback} module shifts the participant's role from "interviewee" to "advisor" and "mentor," allowing them to share wisdom, which enhances their sense of value and meaning.

\subsection{Data Analysis Method}
We employed Braun \& Clarke's six-step thematic analysis method. The steps included: (1) \textbf{Familiarization with data}: Researchers repeatedly read the transcripts. (2) \textbf{Generating initial codes}: Meaningful segments of the text were labeled (e.g., "desire to be heard," "cherishing memories," "aversion to mechanical sounds"). (3) \textbf{Searching for themes}: Codes were grouped to form initial themes (e.g., "The need for connection: from functional to existential," "The desire for non-judgmental listening"). (4) \textbf{Reviewing themes}: Themes were checked against the codes and the entire dataset, and were then merged, split, or discarded as needed. (5) \textbf{Defining and naming themes}: The core idea of each theme was clearly refined. (6) \textbf{Producing the report}: The analysis was written up, using vivid quotes from the interviews to support each theme.

\subsection{Findings}
We recruited 14 participants, aged 29-63 (M=43.5, SD=11.93), with 8 males and 6 females. They had various illnesses including Advanced Chronic Obstructive Pulmonary Disease (COPD), Amyotrophic Lateral Sclerosis (ALS), and Multiple Sclerosis (MS). Participant details are in Table \ref{tab:participant_info}. We categorized participant needs into four main themes (see Appendix C), which are detailed below.

\begin{table}[h!]
\centering
    \caption{Participant Demographics}
\label{tab:participant_info}
\begin{tabular}{|c|c|c|c|c|p{4.5cm}|}
    \hline
    \textbf{ID} & \textbf{Gender} & \textbf{Age} & \textbf{Urban/Rural} & \textbf{Education} & \textbf{Disease} \\
    \hline
        1 & F & 63 & U & High School & Advanced chronic obstructive pulmonary disease (COPD) \\
        2 & M & 44 & U & Primary School & Amyotrophic lateral sclerosis (ALS) \\
        3 & M & 45 & R & High School & Amyotrophic lateral sclerosis (ALS) \\
        4 & M & 25 & R & Bachelor's & Multiple sclerosis (MS) \\
        5 & F & 29 & U & Bachelor's & Late stage of acquired immunodeficiency syndrome (AIDS) \\
        6 & M & 37 & R & High School & End-Stage Renal Disease (ESRD) \\
        7 & M & 37 & U & Bachelor's & Late stage of acquired immunodeficiency syndrome (AIDS) \\
        8 & M & 33 & R & High School & Late stage of acquired immunodeficiency syndrome (AIDS) \\
        9 & F & 62 & U & High School & End-Stage Liver Disease (ESLD) \\
        10 & M & 41 & U & PhD & End-Stage Renal Disease (ESRD) \\
        11 & F & 35 & R & High School & Late stage of acquired immunodeficiency syndrome (AIDS) \\
        12 & M & 47 & R & Primary School & Amyotrophic lateral sclerosis (ALS) \\
        13 & F & 56 & R & High School & Late stage of acquired immunodeficiency syndrome (AIDS) \\
        14 & F & 55 & U & Bachelor's & Amyotrophic lateral sclerosis (ALS) \\
    \hline
\end{tabular}
\end{table}

\subsubsection{Psychological and Emotional Needs}
This was the most significant area, covering themes of combating existential void, seeking meaning, addressing unfinished business, expressing fears, and maintaining dignity.

\textbf{Combating the existential void} was a universal need. This included: (1) A deep craving for sincere dialogue. Participants often interrupted pleasantries, asking, "Can we just talk?" (P2), showing a desire for genuine connection to affirm their existence. (2) An intense desire for touch, such as holding hands or stroking hair. While not always explicitly stated, the frequent mention of words like "hug" and "touch" in their narratives highlighted this need for physical reassurance of their own existence. (3) The creation of small "rituals," like watering a plant daily, to establish order and affirm they were "still here" (P5).

\textbf{Seeking ultimate confirmation of meaning}. This was expressed through: (1) Becoming "storytellers." All participants engaged in long narratives of their lives, re-editing and reinterpreting key moments with each telling, as if to find a coherent thread. P10 repeatedly discussed failing his college entrance exam, but his emotional framing of the event shifted from regret to acceptance, finally seeing the effort itself as a form of happiness. (2) A diversified definition of "legacy," especially among younger participants, who were concerned with their intellectual or spiritual contributions. P4 lamented that his philosophical refutations of Hegel might "be taken to the grave with me," while P5 asked, "Will you remember me?". (3) A need for external validation. Though subtle, many participants would watch the interviewer intently after answering a question, awaiting a response that validated their existence itself, not their status or wealth.

\textbf{Addressing unfinished business} was a concern for all participants. This involved: (1) A desire to act as a "messenger," to convey a final apology, thanks, or declaration of love. All 14 participants mentioned this, with P7 wishing to apologize to a former doctor. (2) Performing symbolic actions, such as tearing up an old letter or organizing a photo album, as external rituals for internal shifts. P14 noted, "As the end nears, you care more about these trivial formalities". (3) A frequent need for "permission to let go," seeking verbal reassurance from loved ones that it was okay to depart.

\textbf{Expressing fundamental fears and receiving "unconditional holding"} was universal. This included: (1) Inquiries about the dying process. The fear was less about what comes after death and more about the loss of dignity during the process—pain, incontinence, and confusion. P5 shared, "What is death really like? A fellow patient told me about a near-death experience... a dark tunnel, which makes me uneasy". (2) A fear of darkness and silence. P10 stated, "I need a small night light... to fight against the absolute nothingness that symbolizes extinction". (3) A need for "unavoiding sincerity." Participants felt more isolated when others feigned optimism or avoided the topic of death. P6 said, "Let's be sincere... the topic you're reluctant to mention, death, is what I'm most interested in".

\textbf{Maintaining dignity and self-identity} was mentioned by 11 participants. This included: (1) Insistence on personal history, correcting misinterpretations of their stories. P7 requested, "Don't call me Professor... I am just myself". (2) Adherence to aesthetic and personal habits, such as keeping themselves neat for the interview, which represented a response to a fundamental need for aesthetic and spiritual well-being. (3) Delineating privacy boundaries. Even in their most vulnerable state, 7 participants expressed a need for private space or information. P1 stated, "I don't want the doctor to ask about things unrelated to my illness. That's my privacy".

\subsubsection{Practical and Control Needs}
\textbf{Maintaining autonomy and a sense of control}. In the face of losing control, participants tried to establish a "minimal order" by: (1) Hyper-detailing choices. P13 noted agonizing over "whether to place the cup on the left or right," as these choices were the last proof of their agency. (2) Exercising power through refusal, such as refusing food or medication, which could be the only domain where they could assert their will. P8 remarked, "I felt the cafeteria lady wasn't friendly to me... I wasn't respected. It's not like I must eat this meal". (3) Creating personal "rules," like the specific arrangement of items, to create a small "kingdom" they still ruled.

\textbf{Arranging posthumous affairs}. This need appeared less frequently than expected, with only 3 participants mentioning it. It included: (1) A pursuit of precision and immutability in instructions for wills and funerals. P9 called it "directing my life's final act". (2) Assigning symbolic items, like a book or piece of jewelry, to specific people, and managing digital legacies like social media accounts, fearing their digital "soul" would become an unmanaged ghost.

\subsubsection{Social and Existential Needs}
\textbf{The desire to be "witnessed" and eternally remembered} to combat "ultimate forgetting" was mentioned by all participants. This involved: (1) Binding their name to their story. P3 said, "You might not remember me, but you'll remember that joke... if you remember me because of it, that would be my honor". (2) Leaving a physical trace, such as a handprint or recording, as "physical evidence that I was here" (P10). (3) Seeking a "custodian of memory," formally asking the interviewer to remember a specific detail for them.

\textbf{Establishing a final, symbolic connection} for a "poetic farewell" with the world. This included: (1) Reconnecting with natural elements, like wanting to see one last sunset or listen to the rain, as a return to the source of life. (2) Achieving transcendence through art, by listening to a piece of music that encapsulates their life or reading a soul-stirring poem. (3) Completing a "narrative loop," such as P2's wish to return to his childhood home, even if physically impossible, to formally close his life's story.

\textbf{Integration and reconciliation}. This included: (1) Forgiving themselves, with some participants acknowledging, "I did my best" (P1). (2) Reframing traumatic narratives, reinterpreting past suffering as a source of strength. (3) Reaching a reconciliation with fate. Instead of anger, there was a calm acceptance, with phrases like "this is my unique life" appearing multiple times.

\subsubsection{Spiritual and Transcendental Needs}
\textbf{Exploring the ultimate meaning of life and death}. This was expressed by: (1) Questioning and reconstructing faith. Both the devout (P9) and atheists (P4) engaged in a process of establishing a more personal, authentic ultimate belief system. (2) Contemplating the vastness of the cosmos. Participants talked about the stars and infinite time, which provided a strange comfort by contextualizing their own mortality as something small (P11). (3) Seeking the "I am" behind "who I am." Six participants described stripping away social roles to explore their most essential core of consciousness.

\textbf{Seeking a sense of connection beyond the individual}. This manifested as: (1) Concern for the future of humanity, extending their care from personal circles to the community and the planet, as a way to transcend their own finitude. (2) A wish to become "nourishment," hoping their experiences could serve as lessons or inspiration for others. (3) Experiencing a "sense of unity," feeling the dissolution of self-boundaries and a merging with all things, which was the ultimate transcendence of loneliness.

\textbf{Experiencing peace and letting go}. This included: (1) A shift from language to silence. After their inner work was done, participants needed a "co-present silence," feeling a peace beyond words. All 14 participants lingered in a quiet, warm, and calm companionship with the interviewer after the formal interview ended. (2) A turning inward of attention, from the external world to inner feelings, as if listening to a call from the depths of life.

\section{Prototype Design}
\subsection{Prototype Design Principles}
Based on the user needs, we proposed a three-pillar design framework for an end-of-life care system (see Appendix D), aimed at addressing the profound psychosocial, existential, and physiological needs of users in this vulnerable stage. The framework emphasizes empathetic interaction, proactive management, and sacred ethics, comprising twelve foundational principles across three core pillars: Interaction Experience, System Functionality, and Technical Ethics.

\subsubsection{Pillar I: Interaction Experience - Humanistic Empathetic Connection}
This pillar governs the user interface and interaction paradigms, prioritizing engagement based on the user's psychological state and declining physical abilities.
\begin{itemize}
    \item \textbf{Principle 1: Presumed Empathy and Emotional Validation.} Rooted in the psychological need for unconditional positive regard, this principle posits that users need understanding, not instruction. Sincere emotional resonance builds trust where platitudes would increase isolation.
    \item \textbf{Principle 2: User Sovereignty and Empowerment.} Stemming from the need to maintain autonomy, this principle dictates that interactional control is key to preserving dignity. Practices include constantly seeking explicit consent and framing interactions as choices, not commands.
    \item \textbf{Principle 3: Seamless Invisibility and Natural Interaction.} Derived from the need to maintain self-identity, this principle argues that users should interact with a "perceived presence," not a machine. Technical visibility can be a reminder of their fragility. This is achieved through minimalist interfaces and natural language dialogue.
    \item \textbf{Principle 4: Hyper-Availability and Inclusive Design.} This directly responds to the user's extreme fatigue. Any operational difficulty can translate into frustration and helplessness. Practices demand minimal interaction latency and support for multiple modalities (voice, light touch, gesture).
\end{itemize}

\subsubsection{Pillar II: System Functionality - Proactive Life Custodian}
This pillar defines the system's core functions, positioning it as an active, guardian-like entity that addresses existential concerns and practical needs.
\begin{itemize}
    \item \textbf{Principle 5: Proactive Care and Initiation.} Grounded in the struggle against existential nihilism, this principle assumes users may lack the energy to initiate contact. The system's proactive engagement demonstrates that the user is still "seen" and valued.
    \item \textbf{Principle 6: Life Narrative and Eternal Witness.} Addressing the profound desire to be witnessed and remembered, the system acts as a final listener and archivist, facilitating a form of "social immortality" through guided life-review dialogues and automated management of a digital legacy.
    \item \textbf{Principle 7: Meaning-Making and Value Affirmation.} Stemming from the ultimate need to confirm life's meaning, the system's dialogue design should promote a positive life review, helping users identify their core values and spiritual legacy.
    \item \textbf{Principle 8: Somatic Comfort and Symptom Management.} Rooted in the fear of the dying process, this principle links physical comfort to psychological peace. Practices include seamless control of environmental systems and providing distraction content to alleviate pain.
    \item \textbf{Principle 9: Social Connection and External Bridging.} Fulfilling the need to establish final, symbolic connections with the outside world, the system must simplify the process of contacting loved ones and intelligently connect users with community resources.
\end{itemize}

\subsubsection{Pillar III: Technical Ethics - Sacred Responsibility and Trust}
Given the user's absolute vulnerability, this pillar elevates technical requirements to the level of moral imperatives.
\begin{itemize}
    \item \textbf{Principle 10: Absolute Reliability and Zero-Failure Mandate.} Derived from the user's profound emotional dependency, a system failure is akin to abandonment by one's sole confidant at a moment of greatest need—a potentially catastrophic event. This requires an extreme pursuit of robustness in system architecture and maintenance.
    \item \textbf{Principle 11: Sacred Data and Legacy Stewardship.} This principle treats the user's digital assets as an extension of their life and final will. A breach of this trust is tantamount to negating the value of their existence. It demands the highest standards of data encryption and, crucially, the faithful execution of posthumous data transfer directives.
    \item \textbf{Principle 12: Privacy Protection and Ethical Primacy.} Rooted in the need to protect the user's final dignity, this principle asserts that the most vulnerable thoughts shared at the end of life must be guarded with the utmost stringency. Practices require transparent and ethical handling of all sensitive data, with special protections for innermost confessions and posthumous arrangements.
\end{itemize}

\subsection{Prototype Design Results}
As shown in Appendix E, we proposed 10 IUI prototype concepts. These systematically construct an intelligent interaction system centered on the end-of-life user, precisely corresponding to their multi-dimensional core needs.

On a \textbf{psychological and emotional level}, the IUI prototypes primarily act as an "indefatigable empathic companion and narrative guardian." The Deep Dialogue Partner and Emotional Soothing Menu address the needs to "combat existential void" and "express fundamental fears." The Life Memoir Creator and Meaning of Life Explorer are dedicated to helping users "seek ultimate confirmation of meaning." The Wishes and Regrets Assistant is designed to "address unfinished business," while the Personalized Environment Steward helps "maintain dignity and self-identity".

On a \textbf{practical and control level}, the IUI prototypes become a "seamless executor of commands and extension of will." The Personalized Environment Steward empowers users to "maintain autonomy and a sense of control," while the Digital Legacy Guide directly serves the need to "arrange posthumous affairs".

On a \textbf{social and existential level}, the prototypes take on the role of an "eternal bridge of connection and memory sculptor." The Life Memoir Creator and Digital Legacy Guide respond to the deep-seated desire to "be witnessed and eternally remembered." The Remote Hug Experience and Virtual Poetic Journey are designed to help users "establish a final symbolic connection." The Wishes and Regrets Assistant also aids in achieving "integration and reconciliation".

Finally, on a \textbf{spiritual and transcendental level}, the IUI acts as a "partner in transcendent dialogue and guide to tranquility." The Meaning of Life Explorer accompanies users to "explore the ultimate meaning of life and death," while the Virtual Poetic Journey and Serene Moments Guide support users to "experience peace and letting go".

All IUI prototype designs include explicit precautions, reflecting careful consideration of technical boundaries, user safety, and privacy, ensuring the appropriateness and reliability of the interface in undertaking such a profound and sacred mission.

\section{Prototype Evaluation}
\subsection{Method}
We re-contacted the participants from section 3.1, along with their doctors and nurses, for a second visit to evaluate our prototype concepts. By this time, participant P9 had passed away. All other participants were re-assessed by their doctors as physically and mentally fit to participate again (using the same criteria as in 3.1). We asked participants to evaluate the 10 prototypes using a Likert scale and to provide suggestions. The evaluation questionnaire is in Appendix F. Ethical assurances and interview procedures were identical to the initial phase.

\subsection{Evaluation Results}
\begin{table}[h!]
\centering
\caption{Participant Ratings of IUI Prototypes}
\label{tab:prototype_ratings}
\begin{tabular}{|c|ccccccccccccc|}
    \hline
    \textbf{IUI} & \textbf{P1} & \textbf{P2} & \textbf{P3} & \textbf{P4} & \textbf{P5} & \textbf{P6} & \textbf{P7} & \textbf{P8} & \textbf{P10} & \textbf{P11} & \textbf{P12} & \textbf{P13} & \textbf{P14} \\
    \hline
    1 & 4 & 5 & 5 & 5 & 5 & 5 & 5 & 5 & 5 & 5 & 5 & 5 & 5 \\
    2 & 4 & 4 & 5 & 5 & 5 & 5 & 5 & 5 & 5 & 5 & 5 & 4 & 4 \\
    3 & 4 & 4 & 5 & 5 & 5 & 5 & 4 & 5 & 5 & 5 & 5 & 4 & 4 \\
    4 & 4 & 4 & 4 & 5 & 4 & 5 & 4 & 5 & 4 & 5 & 5 & 4 & 4 \\
    5 & 4 & 4 & 5 & 4 & 4 & 5 & 5 & 5 & 4 & 5 & 4 & 4 & 4 \\
    6 & 4 & 4 & 5 & 5 & 4 & 5 & 4 & 5 & 5 & 4 & 5 & 4 & 4 \\
    7 & 4 & 4 & 5 & 5 & 5 & 5 & 5 & 5 & 5 & 5 & 5 & 4 & 4 \\
    8 & 5 & 5 & 5 & 5 & 5 & 5 & 5 & 5 & 5 & 5 & 5 & 5 & 5 \\
    9 & 5 & 5 & 5 & 5 & 5 & 5 & 5 & 5 & 5 & 5 & 5 & 5 & 5 \\
    10 & 4 & 4 & 5 & 5 & 5 & 5 & 5 & 4 & 5 & 5 & 5 & 5 & 5 \\
    \hline
    \textbf{Favorite} & 8 & 8 & 8 & 9 & 8 & 8 & 1 & 1 & 8 & 8 & 1 & 8 & 8 \\
    \textbf{Least Fav.} & 2 & 2 & 6 & 2 & 5 & 2 & 2 & 2 & 2 & 2 & 2 & 2 & 2 \\
    \hline
\end{tabular}
\caption*{Note: Ratings are on a 1-5 scale. P9 was deceased at the time of evaluation.}
\end{table}

\textbf{IUI1 \& IUI9: The Boundary between "Tool" and "Companion."} Participants were positive about these dialogue-based IUIs. The value of IUI1 as an "untiring listener" was highly praised. P2 said, "I could tell it the fragments in my mind at any time, without worrying if it would get bored or sad," highlighting its unique value as an emotional buffer. For IUI9, participants appreciated its stance of not providing standard answers. P7 noted, "You said it wouldn't try to give me an answer, but rather, like a knowledgeable librarian, open different windows for me to see for myself. The process itself is more important than getting a conclusion".

\textbf{IUI3 \& IUI4: "Editing Tools" for Reconstructing Life Narratives.} These functions were seen as powerful "narrative tools." IUI3's value was in structuring scattered memories. P1 commented, "The timeline and themes the AI generated for me were like stringing scattered pearls into a necklace". IUI4 was called a "bridge from emotion to action." P5 stated, "Maybe IUI4 can help me draft dialogues for things that are hard for me to say, like apologizing... it turns a huge mountain into steps you can climb".

\textbf{IUI5 \& IUI10: The Experience of Being "Seen" and "Respected."} The value of these immediate psychological comfort functions lay not just in their content but their interaction mode. Many noted the "menu" format of IUI5 was itself therapeutic. P11 explained, "If the AI says, 'It sounds like you're feeling uneasy. Should we talk or listen to some music?' I feel my emotions are seen, and I'm given the right to choose how to be comforted. In an environment where physical autonomy is constantly lost, this micro-level choice is crucial". IUI10's "guardian mode" was also approved. P12 said, "If the AI silently guards me after I fall asleep... it creates a feeling of being safely enveloped. It’s less like a tool and more like a guardian spirit who understands you".

\textbf{IUI6 \& IUI7: Maintaining the "Self" and Continuing "Will."} For IUI6, participants saw it as the last bastion for maintaining selfhood in an uncontrollable environment. P6 said, "Perhaps when the room's light, temperature, and music are still set to 'my' preferences, I am still 'me'... It's helping me defend the last territory of my personality". IUI7 touched upon the deep need for "social immortality." P11 emotionally stated, "The process of organizing my digital legacy is like preparing a home for my soul. Ensuring these photos and texts that carry the joys and sorrows of my life go where they should brings me peace. My story won't completely end with my departure".

\textbf{IUI8 \& IUI2: A Stark Divergence.} This split reflects users' deep thinking on the core issue of "authenticity versus symbolism" at the end of life. IUI8 (Virtual Poetic Journey) received the highest preference. Positive evaluations focused on three dimensions: (1) "The liberation of the soul" was a recurring theme. P2 described, "I look forward to the VR taking me back to the rice fields of my childhood home; I feel like I'm breathing the air of my childhood again. This experience can transcend the confines of the hospital bed and give me one last taste of freedom". (2) The "poetic farewell ceremony" met the aesthetic needs of users for the end of life. (3) Notably, participants felt this experience was "more profound than a real alternative." P11 explained, "A real trip would be limited by my physical condition, but this virtual experience can provide the most profound spiritual journey at my most vulnerable moment".

In contrast, IUI2 (Remote Hug Experience) sparked a profound discussion on the "boundary between technology and humanity." The core controversy was the "ethical dilemma of simulated emotion". P7 pointedly remarked, "This hug is a programmed heartbeat, a preset temperature. It actually deepens my sense of loneliness. It's like a reminder that a real hug is no longer attainable". More importantly, participants were wary of "technology replacing real human connection." P1 reflected, "I worry this technology might make the relatives of those with family feel they've done what's needed, thus reducing actual visits. I think some things shouldn't be simulated".

This divergence reveals a core expectation for IUIs at the end of life: technology should not attempt to "replace" real human connection but should provide unique value that "humans cannot provide." The success of IUI8 lies in creating a transcendent experience beyond human capability, while the controversy of IUI2 stems from its attempt to simulate a basic emotional connection that humans are meant to provide.

\section{Discussion}
This study systematically revealed the complex needs of lonely, terminally ill individuals across four dimensions—psychological-emotional, practical-control, social-existential, and spiritual-transcendent—and designed a set of highly targeted IUI prototypes based on these findings. The evaluation results showed significant differences in acceptance among the prototypes, providing crucial insights into the appropriate role and boundaries of technology at the end of life.

\subsection{Interpretation of Core Findings}
First, the high preference for IUI8 (Virtual Poetic Journey) and the major controversy over IUI2 (Remote Hug Experience) point to a core design principle: in the context of end-of-life care, the value of technology lies in creating "experiences that humans cannot provide," not in "simulating basic connections that humans should provide". IUI8's success was in using VR to transcend physical limitations, enabling a liberation of the spirit. In contrast, IUI2's attempt to technologically simulate a human hug reminded users of the absence of genuine connection, touching upon the ethical dilemma of "simulated emotion". This confirms the concern of P1: "Some things shouldn't be simulated".

Second, the needs map revealed by this study indicates that the design of end-of-life care IUIs must shift from a "functional tool" to an "existential partner" paradigm. While traditional medical tech focuses on physiological monitoring (functional), the high regard for IUI1 (Deep Dialogue Partner), IUI3 (Life Memoir Creator), and IUI9 (Meaning of Life Explorer) shows that users' core expectation is for technology to support life narrative integration, meaning-making, and existential dialogue (existential).

Third, the positive feedback on IUI5 (Emotional Soothing Menu) and IUI6 (Personalized Environment Steward) highlights the extreme importance of a "sense of micro-control" in an environment of macro-level loss of control. Providing "selectable" comfort options and maintaining personal habits are the last bastions for users to defend their dignity and self-identity. This demonstrates that IUIs can effectively alleviate feelings of helplessness by granting users fine-grained control over their environment and interaction flows.

\subsection{Theoretical and Practical Implications}
Theoretically, the proposed "Three Pillars, Twelve Principles" framework elevates technical ethics to a level of importance equal to interaction experience and system functionality, providing a solid theoretical supplement for HCI research with extremely vulnerable groups. Principles like "Sacred Data and Legacy Stewardship" and "Absolute Reliability and Zero-Failure Mandate" emphasize that a single technical failure is no longer a simple tech issue but a betrayal of the user's final trust, with potentially devastating psychological consequences.

Practically, this research offers concrete pathways for social support and technological intervention in end-of-life care. These IUI prototypes can act as "social support amplifiers," filling emotional gaps and helping users complete their final affairs with dignity and order through features like the "Digital Legacy Guide," thus fulfilling the desire for "social immortality".

\subsection{Limitations and Future Work}
This study has limitations. First, while the sample reached saturation for a qualitative study, it was relatively concentrated. Future research could expand the sample range and types of illnesses to test the universality of the design principles. Second, this study primarily evaluated attitudes toward prototype concepts. The next stage requires developing high-fidelity functional prototypes for long-term field deployment to observe real-world effects, challenges, and long-term impacts.

\section{Conclusion}
This study systematically identified the complex needs of lonely, terminally ill individuals across four dimensions—psychological-emotional, practical-control, social-existential, and spiritual-transcendent—and subsequently designed and evaluated a set of targeted IUI prototypes. The core conclusions are as follows:

The Guiding Principle is "Transcendence Over Simulation": A stark divergence in user acceptance revealed a key design insight. The "Virtual Poetic Journey" (IUI8) was highly favored for its ability to use technology to transcend physical limitations and enable a liberation of the spirit. Conversely, the "Remote Hug Experience" (IUI2) was controversial because its attempt to simulate a fundamental human connection served as a painful reminder of the absence of genuine contact, touching upon the ethical dilemma of "simulated emotion". This establishes a critical directive: the value of technology in end-of-life care lies in creating experiences that humans cannot provide, not in simulating basic connections that they should. A Paradigm Shift from "Functional Tool" to "Existential Partner" is Necessary: The findings indicate that users' primary expectations of technology are to support life narrative integration, meaning-making, and existential dialogue, rather than merely performing functional tasks like physiological monitoring. This requires a design philosophy that conceives of the IUI as an active "existential partner" rather than a passive "functional tool".

"Micro-Control" is Crucial for Preserving Dignity: In an environment where individuals experience a macro-level loss of control, providing fine-grained, selectable "micro-control" over their personal environment and interaction flows via an IUI is a vital last bastion for defending their dignity and self-identity. Technical Ethics are as Important as Functionality and Experience: The proposed "Three Pillars, Twelve Principles" framework elevates technical ethics, with principles like "Sacred Data and Legacy Stewardship" and an "Absolute Reliability and Zero-Failure Mandate," to a level of importance equal to system functionality and interaction experience. This underscores that for this extremely vulnerable group, a technical failure is not a simple glitch but a potentially devastating psychological betrayal.

In summary, this research offers a clear, critical, and user-centered direction for HCI in end-of-life care, advocating a shift from designing "tools for dying" toward designing "partners for living with dignity until the very end," thereby meaningfully addressing the ultimate challenge of loneliness at the end of life.

\bibliographystyle{ACM-Reference-Format}
\bibliography{sample-base}

@inproceedings{wallace2020hci,
  title={HCI at end of life \& beyond},
  author={Wallace, Jayne and Odom, Will and Montague, Kyle and Koulidou, Nantia and Sas, Corina and Morrissey, Kellie and Olivier, Patrick},
  booktitle={Extended Abstracts of the 2020 CHI Conference on Human Factors in Computing Systems},
  pages={1--8},
  year={2020}
}

@inproceedings{ahmadpour2023can,
  title={How can HCI support end-of-life care? Critical perspectives on sociotechnical imaginaries for palliative care},
  author={Ahmadpour, Naseem and Gough, Phillip and Lovell, Melanie and Austin, Philip and Poronnik, Philip and Zhang, Wendy Qi and Kay, Judy and Kummerfeld, Bob and Luckett, Tim and Brown, Martin and others},
  booktitle={Extended Abstracts of the 2023 CHI Conference on Human Factors in Computing Systems},
  pages={1--7},
  year={2023}
}

@inproceedings{massimi2011matters,
  title={Matters of life and death: locating the end of life in lifespan-oriented HCI research},
  author={Massimi, Michael and Odom, William and Banks, Richard and Kirk, David},
  booktitle={Proceedings of the sigchi conference on human factors in computing systems},
  pages={987--996},
  year={2011}
}

@inproceedings{albers2023dying,
  title={Dying, Death, and the Afterlife in Human-Computer Interaction. A Scoping Review.},
  author={Albers, Ruben and Sadeghian, Shadan and Laschke, Matthias and Hassenzahl, Marc},
  booktitle={Proceedings of the 2023 CHI Conference on Human Factors in Computing Systems},
  pages={1--16},
  year={2023}
}

@article{plotzke2021construction,
  title={Construction and Performance of the Hospice Care Index Claims-Based Quality Measure},
  author={Plotzke, Michael and Christian, Thomas and Groover, Kim and Harrison, Zinnia and Abdur-Rahman, Ihsan and Massuda, Cindy},
  journal={Innovation in Aging},
  volume={5},
  number={Suppl 1},
  pages={62},
  year={2021}
}

@inproceedings{saito2025unintended,
  title={Unintended, Percolated Work: Overlooked Opportunities for Collaboration Between Informal Caregivers and Healthcare Professionals During the End-Of-Life Care Process},
  author={Saito, Shun and Sugihara, Taro},
  booktitle={Proceedings of the 2025 CHI Conference on Human Factors in Computing Systems},
  pages={1--16},
  year={2025}
}

@inproceedings{chen2021happens,
  title={What happens after death? Using a design workbook to understand user expectations for preparing their data},
  author={Chen, Janet X and Vitale, Francesco and McGrenere, Joanna},
  booktitle={Proceedings of the 2021 CHI Conference on Human Factors in Computing Systems},
  pages={1--13},
  year={2021}
}

@inproceedings{massimi2013exploring,
  title={Exploring remembrance and social support behavior in an online bereavement support group},
  author={Massimi, Michael},
  booktitle={Proceedings of the 2013 conference on Computer supported cooperative work},
  pages={1169--1180},
  year={2013}
}

@article{kuziemsky2006hospice,
  title={The e-Hospice—Beyond traditional boundaries of palliative care},
  author={Kuziemsky, Craig E and Jahnke, Jens H and Lau, Francis},
  journal={Telematics and Informatics},
  volume={23},
  number={2},
  pages={117--133},
  year={2006},
  publisher={Elsevier}
}

@inproceedings{massimi2011dealing,
  title={Dealing with death in design: developing systems for the bereaved},
  author={Massimi, Michael and Baecker, Ronald M},
  booktitle={Proceedings of the SIGCHI Conference on Human Factors in Computing Systems},
  pages={1001--1010},
  year={2011}
}

@inproceedings{kim2024maintaining,
  title={Maintaining Continuing Bonds in Bereavement: A Participatory Design Process of Be. side},
  author={Kim, Jieun and Uriu, Daisuke and Barbareschi, Giulia and Kamiyama, Youichi and Minamizawa, Kouta},
  booktitle={Proceedings of the 2024 CHI Conference on Human Factors in Computing Systems},
  pages={1--15},
  year={2024}
}

@inproceedings{odom2010passing,
  title={Passing on \& putting to rest: understanding bereavement in the context of interactive technologies},
  author={Odom, William and Harper, Richard and Sellen, Abigail and Kirk, David and Banks, Richard},
  booktitle={Proceedings of the SIGCHI conference on Human Factors in computing systems},
  pages={1831--1840},
  year={2010}
}

@inproceedings{beijer2025mano,
  title={Mano: Designing for Tactile Experiences in Advanced Dementia Care},
  author={Beijer, Sanne and Houben, Maarten and IJsselsteijn, Wijnand and Brankaert, Rens},
  booktitle={Proceedings of the 2025 CHI Conference on Human Factors in Computing Systems},
  pages={1--13},
  year={2025}
}

@inproceedings{ullal2024iterative,
  title={An iterative participatory design approach to develop collaborative augmented reality activities for older adults in long-term care facilities},
  author={Ullal, Akshith and Tauseef, Mahrukh and Watkins, Alexandra and Juckett, Lisa and Maxwell, Cathy A and Tate, Judith and Mion, Lorraine and Sarkar, Nilanjan},
  booktitle={Proceedings of the 2024 CHI Conference on Human Factors in Computing Systems},
  pages={1--21},
  year={2024}
}

@inproceedings{ferguson2014craving,
  title={Craving, creating, and constructing comfort: insights and opportunities for technology in hospice},
  author={Ferguson, Robert Douglas and Massimi, Michael and Crist, Emily Anne and Moffatt, Karyn Anne},
  booktitle={Proceedings of the 17th ACM conference on Computer supported cooperative work \& social computing},
  pages={1479--1490},
  year={2014}
}

@article{abdel2022development,
  title={Development of a user-centered design framework for palliative and hospice care patients for a better quality of life experience},
  author={Abdel-Razek, Shahira Assem},
  journal={Environmental Science \& Sustainable Development},
  volume={7},
  number={1},
  pages={105--117},
  year={2022}
}

@article{cox2018palliative,
  title={Palliative care planner: a pilot study to evaluate acceptability and usability of an electronic health records system-integrated, needs-targeted app platform},
  author={Cox, Christopher E and Jones, Derek M and Reagan, Wen and Key, Mary D and Chow, Vinca and McFarlin, Jessica and Casarett, David and Creutzfeldt, Claire J and Docherty, Sharron L},
  journal={Annals of the American Thoracic Society},
  volume={15},
  number={1},
  pages={59--68},
  year={2018},
  publisher={American Thoracic Society}
}

@article{jeon2025targeted,
  title={Targeted Digital Health Intervention in End-of-Life and Hospice Care: A Scoping Review},
  author={Jeon, Misun and Jeon, Heejung and Kim, Sanghee},
  journal={Journal of advanced nursing},
  year={2025},
  publisher={Wiley Online Library}
}

@inproceedings{kim2023alphadapr,
  title={Alphadapr: An ai-based explainable expert support system for art therapy},
  author={Kim, Jiwon and Kang, Jiwon and Kim, Taeeun and Song, Hayeon and Han, Jinyoung},
  booktitle={Proceedings of the 28th International Conference on Intelligent User Interfaces},
  pages={19--31},
  year={2023}
}

@article{murphree2021improving,
  title={Improving the delivery of palliative care through predictive modeling and healthcare informatics},
  author={Murphree, Dennis H and Wilson, Patrick M and Asai, Shusaku W and Quest, Daniel J and Lin, Yaxiong and Mukherjee, Piyush and Chhugani, Nirmal and Strand, Jacob J and Demuth, Gabriel and Mead, David and others},
  journal={Journal of the American Medical Informatics Association},
  volume={28},
  number={6},
  pages={1065--1073},
  year={2021},
  publisher={Oxford Academic}
}

@article{tan2024scoping,
  title={A scoping review of digital technology applications in palliative care},
  author={Tan, YinHu and Liang, Xue and Ming, Wei and Xing, HuiMin and Wang, Yang and Gao, Yan},
  journal={BMC Palliative Care},
  volume={23},
  number={1},
  pages={290},
  year={2024},
  publisher={Springer}
}

@article{carey2023co,
  title={Co-design and prototype development of the ‘Ayzot App’: A mobile phone based remote monitoring system for palliative care},
  author={Carey, Nicola and Abathun, Ephrem and Maguire, Roma and Wodaje, Yohans and Royce, Catherine and Ayers, Nicola},
  journal={Palliative Medicine},
  volume={37},
  number={5},
  pages={771--781},
  year={2023},
  publisher={SAGE Publications Sage UK: London, England}
}

@article{courtright2019electronic,
  title={Electronic health record mortality prediction model for targeted palliative care among hospitalized medical patients: a pilot quasi-experimental study},
  author={Courtright, Katherine R and Chivers, Corey and Becker, Michael and Regli, Susan H and Pepper, Linnea C and Draugelis, Michael E and O’Connor, Nina R},
  journal={Journal of general internal medicine},
  volume={34},
  number={9},
  pages={1841--1847},
  year={2019},
  publisher={Springer}
}

@article{blanes2023user,
  title={User-centred design of a clinical decision support system for palliative care: Insights from healthcare professionals},
  author={Blanes-Selva, Vicent and Asensio-Cuesta, Sabina and Do{\~n}ate-Mart{\'\i}nez, Ascensi{\'o}n and Pereira Mesquita, Felipe and Garc{\'\i}a-G{\'o}mez, Juan M},
  journal={Digital Health},
  volume={9},
  pages={20552076221150735},
  year={2023},
  publisher={SAGE Publications Sage UK: London, England}
}

@article{maguraushe2024use,
  title={The use of smart technologies for enhancing palliative care: A systematic review},
  author={Maguraushe, Kudakwashe and Ndlovu, Belinda Mutunhu},
  journal={Digital health},
  volume={10},
  pages={20552076241271835},
  year={2024},
  publisher={SAGE Publications Sage UK: London, England}
}

@article{sandham2022intelligent,
  title={Intelligent palliative care based on patient-reported outcome measures},
  author={Sandham, Margaret H and Hedgecock, Emma A and Siegert, Richard J and Narayanan, Ajit and Hocaoglu, Mevhibe B and Higginson, Irene J},
  journal={Journal of pain and symptom management},
  volume={63},
  number={5},
  pages={747--757},
  year={2022},
  publisher={Elsevier}
}

@inproceedings{doganguen2025speech,
  title={Speech-Based Assistants in Professional Healthcare: Potentials and Challenges in Palliative Care},
  author={Doganguen, Ayseguel and Oelcer, Sabahat},
  booktitle={International Conference on Human-Computer Interaction},
  pages={238--251},
  year={2025},
  organization={Springer}
}

@article{narvaez2025artificial,
  title={Artificial intelligence in symptom management and clinical decision support for palliative care},
  author={Narvaez, Roison Andro and Ferrer, Marilane and Peco, Ralph Antonio and Mejilla, Joylyn},
  journal={International Journal of Palliative Nursing},
  volume={31},
  number={6},
  pages={294--306},
  year={2025},
  publisher={MA Healthcare London}
}

@inproceedings{deng2025research,
  title={Research on Design Strategy of Home Palliative Care Virtual Reality Service Based on SAPAD-AHP-SUS},
  author={Deng, Yun and Tang, Shichen and Gao, Jie},
  booktitle={International Conference on Human-Computer Interaction},
  pages={18--37},
  year={2025},
  organization={Springer}
}

@article{abernethy2010strategy,
  title={A strategy to advance the evidence base in palliative medicine: formation of a palliative care research cooperative group},
  author={Abernethy, Amy P and Aziz, Noreen M and Basch, Ethan and Bull, Janet and Cleeland, Charles S and Currow, David C and Fairclough, Diane and Hanson, Laura and Hauser, Joshua and Ko, Danielle and others},
  journal={Journal of palliative medicine},
  volume={13},
  number={12},
  pages={1407--1413},
  year={2010},
  publisher={Mary Ann Liebert, Inc. 140 Huguenot Street, 3rd Floor New Rochelle, NY 10801 USA}
}

@article{ritchie2017better,
  title={Better together: The making and maturation of the palliative care research cooperative group},
  author={Ritchie, Christine L and Pollak, Kathryn I and Kehl, Karen A and Miller, Jeri L and Kutner, Jean S},
  journal={Journal of Palliative Medicine},
  volume={20},
  number={6},
  pages={584--591},
  year={2017},
  publisher={Mary Ann Liebert, Inc. 140 Huguenot Street, 3rd Floor New Rochelle, NY 10801 USA}
}

@article{suslow2023impact,
  title={Impact of information and communication software on multiprofessional team collaboration in outpatient palliative care--a qualitative study on providers’ perspectives},
  author={Suslow, Anastasia and Giehl, Chantal and Hergesell, Jannis and Vollmar, Horst Christian and Otte, Ina},
  journal={BMC Palliative Care},
  volume={22},
  number={1},
  pages={19},
  year={2023},
  publisher={Springer}
}

@article{kuziemsky2008interdisciplinary,
  title={An interdisciplinary computer-based information tool for palliative severe pain management},
  author={Kuziemsky, Craig E and Weber-Jahnke, Jens H and Lau, Francis and Downing, G Michael},
  journal={Journal of the American Medical Informatics Association},
  volume={15},
  number={3},
  pages={374--382},
  year={2008},
  publisher={BMJ Group BMA House, Tavistock Square, London, WC1H 9JR}
}

@inproceedings{smith2024designing,
  title={(Un) designing AI for Mental and Spiritual Wellbeing},
  author={Smith, C Estelle and Bezabih, Alemitu and Freed, Diana and Halperin, Brett A and Wolf, Sara and Claisse, Caroline and Li, Jingjin and Hoefer, Michael and Rifat, Mohammad Rashidujjaman},
  booktitle={Companion Publication of the 2024 Conference on Computer-Supported Cooperative Work and Social Computing},
  pages={117--120},
  year={2024}
}

@inproceedings{wu2024collective,
  title={Collective imaginaries for the futures of care work},
  author={Wu, Yiying and Lee, Jung-Joo and Pillai, Ajit G and Cho, Janghee and Ahmadpour, Naseem and Roto, Virpi and Sachathep, Thida and Liu, Jiashuo and Sawan, Mouna and Song, Dongjin and others},
  booktitle={Companion Publication of the 2024 Conference on Computer-Supported Cooperative Work and Social Computing},
  pages={732--735},
  year={2024}
}

@article{Ostherr2016,
  title = {Death in the Digital Age: A Systematic Review of Information and Communication Technologies in End-of-Life Care},
  author = {Kirsten Ostherr and Peter Killoran and Christine Shegog and Susan J. Bruera},
  journal = {Journal of Palliative Medicine},
  year = {2016},
  doi = {10.1089/jpm.2015.0341},
  url = {https://doi.org/10.1089/jpm.2015.0341}
}

@inproceedings{MassimiBaecker2011,
  title = {Dealing with Death in Design: Developing Systems for the Bereaved},
  author = {Michael Massimi and Ronald M. Baecker},
  booktitle = {ACM CHI Conference},
  year = {2011},
  doi = {10.1145/1978942.1979092},
  url = {https://doi.org/10.1145/1978942.1979092}
}

@article{Byock2025,
  title = {A Strategic Path Forward for Hospice and Palliative Care},
  author = {Ira Byock},
  journal = {Palliative Medicine Reports},
  year = {2025},
  doi = {10.1089/pmr.2025.0030},
  url = {https://doi.org/10.1089/pmr.2025.0030}
}

@inproceedings{MassimiBaecker2010,
  title = {A Death in the Family: Opportunities for Designing Technologies for the Bereaved},
  author = {Michael Massimi and Ronald M. Baecker},
  booktitle = {ACM CHI Conference},
  year = {2010},
  doi = {10.1145/1753326.1753600},
  url = {https://doi.org/10.1145/1753326.1753600}
}

@article{Abejas2025,
  title = {Ethical Challenges and Opportunities of AI in End-of-Life Palliative Care: Integrative Review},
  author = {Abel García Abejas and David Geraldes Santos and Fabio Leite Costa and others},
  journal = {Interactive Journal of Medical Research},
  year = {2025},
  doi = {10.2196/73517},
  url = {https://doi.org/10.2196/73517}
}

@article{Stanley2024,
  title = {How can technology be used to support communication in palliative care beyond the covid-19 pandemic: a mixed-methods national survey of palliative care healthcare professionals},
  author = {Sarah Stanley and Anne Finucane and Anthony Thompson and Amara Callistus Nwosu},
  journal = {BMC Palliative Care},
  year = {2024},
  doi = {10.1186/s12904-024-01372-z},
  url = {https://doi.org/10.1186/s12904-024-01372-z}
}

@inproceedings{MassimiCharise2009,
  title = {Dying, death, and mortality: towards thanatosensitivity in HCI},
  author = {Michael Massimi and Andrea Charise},
  booktitle = {ACM CHI Extended Abstracts},
  year = {2009},
  doi = {10.1145/1520340.1520349},
  url = {https://doi.org/10.1145/1520340.1520349}
}

@article{Vandersman2024,
  title = {'Technology in end-of-life care is very important': the view of nurses regarding technology and end-of-life care},
  author = {P. Vandersman and J. Tieman and others},
  journal = {BMC Nursing},
  year = {2024},
  doi = {10.1186/s12912-024-02475-x},
  url = {https://doi.org/10.1186/s12912-024-02475-x}
}

@inproceedings{Massimi2010,
  title = {HCI at the end of life: understanding death, dying, and the digital},
  author = {Michael Massimi and William Odom and David Kirk and Richard Banks},
  booktitle = {ACM CHI Extended Abstracts},
  year = {2010},
  doi = {10.1145/1753846.1754178},
  url = {https://doi.org/10.1145/1753846.1754178}
}

\appendix
\section{Interviewer Training and Principles for Research with Lonely, End-of-Life Individuals}
\begin{longtable}{@{}p{0.15\linewidth} p{0.2\linewidth} p{0.55\linewidth}@{}}
    \caption{Interviewer Training and Principles} \label{tab:interviewer_training} \\
    \toprule
    \textbf{Dimension} & \textbf{Module} & \textbf{Specific Content and Requirements} \\
    \midrule
    \endfirsthead
    \multicolumn{3}{c}%
    {{\bfseries Table \thetable\ continued from previous page}} \\
    \toprule
    \textbf{Dimension} & \textbf{Module} & \textbf{Specific Content and Requirements} \\
    \midrule
    \endhead
    \bottomrule
    \multicolumn{3}{r}{{Continued on next page}} \\
    \endfoot
    \bottomrule
    \endlastfoot
    \multirow{4}{=}{Core Theory Training} & Palliative Care Communication Skills & Learn the specifics, principles, and taboos of communicating with the terminally ill (e.g., no false reassurances, not avoiding the topic of death). Master sensitive strategies for delivering bad news. Train on responding to ultimate questions about life's meaning and fear of death. \\
    \cmidrule(lr){2-3}
    & End-of-Life Care Knowledge & Understand the physical, psychological, social, and spiritual needs (total pain) of the terminally ill. Be familiar with common end-stage symptoms (e.g., pain, fatigue) and their impact on communication. Understand the basic concepts of psychological support frameworks like "Dignity Therapy" and "Life Review". \\
    \cmidrule(lr){2-3}
    & Grief Counseling Principles & Understand various grief reactions caused by loss (of bodily function, social roles, the future). Master the core principle of "accompanying grief" rather than "solving grief". \\
    \cmidrule(lr){2-3}
    & In-depth Research Ethics Seminar & Deeply study the three principles of the Belmont Report (Respect for Persons, Beneficence, Justice). Through case studies and role-playing, master the specific practices for protecting "vulnerable populations". \\
    \midrule
    \multirow{5}{=}{Core Skills Training} & Active Listening Techniques & Full engagement: maintain open posture, eye contact, nodding. Listen for emotions and metaphors: capture "feeling words" and analogies (e.g., "like being trapped in a cage"). Embrace silence: allow and respect pauses for thought. \\
    \cmidrule(lr){2-3}
    & Deep Empathy and Emotional Validation & Core phrases: "It sounds like that made you feel very \underline{\hspace{2cm}}." "Thank you for being willing to share that with me." Strictly avoid: empty reassurances ("don't be sad"), negating feelings ("you're overthinking it"), sharing personal experiences. \\
    \cmidrule(lr){2-3}
    & Non-directive Questioning Techniques & Use open-ended questions: "Can you tell me more about how you felt then?" "What did that mean to you?" Avoid leading questions: Strictly forbid questions like "You must have been happy then, right?" Use follow-ups based on their answers to show genuine interest. \\
    \cmidrule(lr){2-3}
    & Crisis Recognition and Intervention & Identify signs of acute emotional distress (agitation, uncontrollable crying, hopeless statements). Standard interruption script: "You seem very upset. Do we need to pause? Would you like me to ask the psychology colleague to come in?" Clear reporting and referral process for emergencies. \\
    \cmidrule(lr){2-3}
    & Self-Care and Resilience Building & Learn to recognize personal emotional exhaustion and "compassion fatigue." Establish regular group supervision and individual counseling mechanisms for emotional support. \\
    \midrule
    \multirow{9}{=}{Mandatory Interview Principles} & Absolute Respect for Autonomy & Must obtain explicit verbal consent for every action ("We're starting the recording now, is that okay?"). Constantly remind participants of their absolute right to "pause, skip, or withdraw at any time". \\
    \cmidrule(lr){2-3}
    & Beneficence and Non-Maleficence & Emotional safety first: if significant emotional distress occurs, questioning must be paused, and support prioritized. Physical comfort first: closely observe signs of fatigue, proactively suggest breaks, and do not exhaust the participant's energy. \\
    \cmidrule(lr){2-3}
    & Strict Confidentiality & Do not discuss participant information in any non-research setting. Ensure all data is protected with the highest level of security during transmission, storage, and processing. \\
    \cmidrule(lr){2-3}
    & "Follow," Don't "Lead" & Let the participant set the pace and depth of the conversation. Allow for "valuable deviations," which often reveal the truest needs. \\
    \cmidrule(lr){2-3}
    & De-expertize & The interviewer's core task is to create a safe space, listen, and witness, not to provide answers, advice, or therapy. For medical questions, guide them to communicate with healthcare staff. \\
    \cmidrule(lr){2-3}
    & Sincerity and Transparency & Do not feign emotions. Can respond sincerely: "Your story has deeply moved me." Acknowledge limitations: "I may not be able to fully understand, but I am very willing to listen". \\
    \cmidrule(lr){2-3}
    & Preparation and Presence & Rest fully before the interview to ensure being "fully present" mentally and physically. During the interview, set aside distractions and give full attention to the participant. \\
    \cmidrule(lr){2-3}
    & Non-Judgment & Maintain complete neutrality and a non-judgmental attitude towards the participant's life choices, values, and emotional reactions. \\
    \cmidrule(lr){2-3}
    & Clear Boundaries & Maintain a professional relationship; do not develop personal friendships. Do not make promises outside the scope of the research that cannot be fulfilled. \\
\end{longtable}

\section{In-Depth Psychological Needs Interview Questionnaire for End-of-Life Lonely Patients}

\begin{longtable}{@{} p{0.12\linewidth} p{0.53\linewidth} p{0.28\linewidth} @{}}
    
    \caption{Interview Questionnaire Structure and Content} \label{tab:interview_questionnaire} \\
    \toprule
    \textbf{Module} & \textbf{Interview Question} & \textbf{Rationale and Objective} \\
    \midrule
    \endfirsthead

    % Header for subsequent pages
    \multicolumn{3}{c}%
    {{\bfseries Table \thetable\ continued from previous page}} \\
    \toprule
    \textbf{Module} & \textbf{Interview Question} & \textbf{Rationale and Objective} \\
    \midrule
    \endhead

    % Footer for pages except the last
    \bottomrule
    \multicolumn{3}{r}{{Continued on next page}} \\
    \endfoot

    % Footer for the last page
    \bottomrule
    \endlastfoot

    % --- Content Start ---

    % Opening
    \textbf{Opening Statement} & 
    ``Hello, I truly appreciate your trust and willingness to spend this time with me. My primary purpose today is to listen to your stories, feelings, and thoughts. We thank you for your generosity; your expression will make a significant contribution to our society, especially to human-centered technology design. On behalf of myself and our research team, I thank you in advance.'' \par\vspace{0.5em}
    ``Please note that I will now read your rights again. You may interrupt me at any time: (1) Our conversation is strictly confidential and anonymized. (2) You may view or delete your data. (3) You have full control to skip questions or pause. (4) You may withdraw at any time. (5) You may ask us to leave immediately if uncomfortable. (6) You may ask to reiterate the research purpose. (7) Medical staff are outside if needed.'' \par\vspace{0.5em}
    ``If there are no concerns, please sign to indicate your understanding. Shall we slowly begin, following your pace?'' & 
    \textbf{Establishing a Safety Zone}: Clarify the tone as listening, confidential, and respectful, granting the interviewee a complete sense of control. \\
    \cmidrule(lr){1-3}

    % Module 1
    \textbf{(1) Establishing Empathic Connection} & 
    \textbf{Q1.} If possible, would you be willing to lead me into the corridor of your memories and share one or two of your most treasured life fragments? Regarding childhood, family, work, or any moment that makes your eyes light up. & 
    \textbf{Initiating Life Story}: Starting from positive, personal memories to guide the interviewee to enter the dialogue as a complete ``person'' rather than a ``patient.'' \\
    \addlinespace
    & \textbf{Q2.} In your long life, which identities or roles (e.g., parent, friend, craftsman, teacher) do you cherish the most? What qualities or achievements are you most proud of? & 
    \textbf{Affirming Value and Identity}: Confirming core values and self-identity of a lifetime to enhance the sense of dignity. \\
    \addlinespace
    & \textbf{Q3.} Having experienced so much, what do you consider the most important life wisdom you have learned? If given the chance, what would you like to say to young children? & 
    \textbf{Distilling Life Wisdom}: Guiding life review and meaning integration; experiencing the transmission of knowledge. \\
    \addlinespace
    & \textbf{Q4.} If you used a metaphor (such as a river, a book, a journey) to describe your life, what would it be? Why? & 
    \textbf{Metaphorical Summary}: Prompting holistic reflection on life through a poetic form. \\
    \addlinespace
    & \textbf{Q5.} In your life, was there a decision or event that completely changed your life trajectory like a crossroad? Do you often think back to those ``what might have been'' scenarios? & 
    \textbf{Exploring Turning Points}: Deepening the narrative by touching upon potential regrets and nostalgia to complete a deeper life review. \\
    \cmidrule(lr){1-3}

    % Module 2
    \textbf{(2) Exploring the Inner World} & 
    \textbf{Q6.} How would you describe your predominant mood recently? Is it peaceful, lonely, fearful, relieved, or an indescribably complex emotion? & 
    \textbf{Assessing Emotional State}: Understanding the core emotional tone to pave the way for deeper topics. \\
    \addlinespace
    & \textbf{Q7.} Regarding your current condition, what specifically troubles or frightens you the most? (e.g., pain, loss of dignity, the unknown process, becoming a burden). & 
    \textbf{Probing Specific Fears}: Concretizing abstract fears to understand the most urgent physical and mental distress. \\
    \addlinespace
    & \textbf{Q8.} Beneath these fears, has there been any moment where you felt unusual peace or a trace of warmth? When was that? & 
    \textbf{Excavating Inner Resources}: Guiding attention to positive moments and sources of strength that still exist. \\
    \addlinespace
    & \textbf{Q9.} At this stage of life, what has become most important to you? What still makes you feel the meaning of life? (e.g., a sincere connection, sunlight, inner contemplation). & 
    \textbf{Seeking Current Meaning}: Focusing on the ``here and now'' to find specific meanings that support survival. \\
    \addlinespace
    & \textbf{Q10.} When you hear the word ``lonely,'' what specifically does it mean to you? Is it ``having no one around,'' or ``even with people around, no one cares about my inner self''? & 
    \textbf{Defining Loneliness}: Refining the morphology of loneliness to understand the lack of specific connection. \\
    \addlinespace
    & \textbf{Q11.} At what time of day does loneliness usually strike most intensely? What triggers it? & 
    \textbf{Identifying Triggers}: Associating loneliness with specific contexts to provide timely support. \\
    \addlinespace
    & \textbf{Q12.} Deep down, what form of companionship do you crave the most? Is it silent physical presence (holding hands)? Emotional resonance (empathy)? Or intellectual exchange? & 
    \textbf{Clarifying Connection Needs}: Helping the interviewee express the most desired mode of companionship. \\
    \addlinespace
    & \textbf{Q13.} If you could say one sentence to anyone in the world (living or deceased) right now, what would you say? To whom? & 
    \textbf{Addressing Unspoken Words}: Providing an opportunity for catharsis (apology, gratitude, or farewell). \\
    \addlinespace
    & \textbf{Q14.} When someone visits you, what do you most hope they do? And what do you least hope they do? (e.g., not just talking about illness, avoiding pity). & 
    \textbf{Guiding Visitor Behavior}: Obtaining expectations for social interaction to reduce social pressure. \\
    \addlinespace
    & \textbf{Q15.} Looking back at the relationships in your life, are there any connections you wished were closer but didn't turn out as hoped? & 
    \textbf{Unfinished Relational Business}: Providing direction for the ``Four Tasks of Life'' (Thanking, Apologizing, Loving, Goodbye). \\
    \addlinespace
    & \textbf{Q16.} When you think of the end of life, do you feel regret because you might lack the companionship of certain people? & 
    \textbf{Confronting Core Emotions}: Assessing the deepest regret and sadness caused by loneliness. \\
    \addlinespace
    & \textbf{Q17.} For these regrets, do you currently tend to hope for a chance to make amends, or do you prefer to accept them silently? & 
    \textbf{Understanding Coping Patterns}: Determining whether the tendency is active resolution or passive acceptance. \\
    \addlinespace
    & \textbf{Q18.} In daily life, what small things (e.g., deciding what to eat, music choices) still make you feel ``I am still me'' and that I still have choices? & 
    \textbf{Preserving Micro-Autonomy}: Protecting daily minutiae that maintain self-identity and control. \\
    \addlinespace
    & \textbf{Q19.} Do you have any daily habits or ``rituals'' you wish to be maintained? (e.g., morning reading, meditation). & 
    \textbf{Respecting Personal Rituals}: Maintaining inner order and normalcy through habits. \\
    \addlinespace
    & \textbf{Q20.} In situations where you need care, what makes you feel the most loss of dignity? How can we do better to maintain it? & 
    \textbf{Protecting Dignity}: Directly inquiring about dignity-damaging aspects to seek improvements. \\
    \addlinespace
    & \textbf{Q21.} When your body gradually ceases to obey, how do you view the relationship between the ``real you'' and ``this body''? & 
    \textbf{Mind-Body Philosophy}: Discussing the relationship between self and the decaying body. \\
    \cmidrule(lr){1-3}

    % Module 3
    \textbf{(3) Seeking Solace} & 
    \textbf{Q22.} When feeling painful or low, what brings you the greatest comfort? (e.g., faith, memories, music, nature). & 
    \textbf{Identifying Comfort Sources}: Identifying accessible psychological soothing resources. \\
    \addlinespace
    & \textbf{Q23.} Have you ever felt connected to something grander than yourself (e.g., nature, the universe, divinity)? How does this affect you? & 
    \textbf{Exploring Transcendent Experiences}: Understanding their spiritual world, which may offer ultimate care. \\
    \addlinespace
    & \textbf{Q24.} Regarding what comes after life, do you have any thoughts or beliefs? Do these thoughts make you feel peaceful or otherwise? & 
    \textbf{Discussing Views on Death}: Respecting views on death to address current fear or peace levels. \\
    \addlinespace
    & \textbf{Q25.} Do you feel that the people around you truly understand your inner feelings and needs? How do you wish they could understand you better? & 
    \textbf{Desire for Understanding}: Assessing the support system and obtaining suggestions for communication. \\
    \cmidrule(lr){1-3}

    % Module 4
    \textbf{(4) Specific Wishes} & 
    \textbf{Q26.} Is there a specific taste you want to taste again? A song to hear again? Or a specific atmosphere (e.g., flowers, soft lighting)? & 
    \textbf{Fulfilling Sensory Wishes}: Focusing on sensory, achievable wishes to improve end-of-life quality. \\
    \addlinespace
    & \textbf{Q27.} Is there a small thing you feel ``it would be good if finished''? (e.g., knowing a story ending, finding a new owner for an item). & 
    \textbf{Completing Unfinished Business}: Helping complete feasible wishes for a final sense of control. \\
    \addlinespace
    & \textbf{Q28.} Do you wish for someone to read specific texts to you? Or need help recording some words or a letter? & 
    \textbf{Assisting Expression}: Providing practical help for information input (listening) or output (recording). \\
    \addlinespace
    & \textbf{Q29.} Besides material possessions, what ``intangible legacy'' do you hope to leave behind? Perhaps a spirit, a story, or a shared memory? & 
    \textbf{Defining Spiritual Legacy}: Guiding reflection on the core memory to be left behind. \\
    \cmidrule(lr){1-3}

    % Module 5
    \textbf{(5) Views on IUI} & 
    \textbf{Q30.} Imagine an intelligent device (like a tablet or smart speaker) that interacts with you. Are you curious about such a thing, or do you have concerns? & 
    \textbf{Assessing Tech Acceptance}: Understanding willingness to use and potential barriers. \\
    \addlinespace
    & \textbf{Q31.} When you feel lonely but don't want to disturb others, how could such an interface help you? (e.g., chatting, reading, playing music, contacting relatives). & 
    \textbf{Exploring Functional Needs}: Exploring technology's potential in filling social voids. \\
    \addlinespace
    & \textbf{Q32.} Conversely, in what aspects do you feel it cannot replace human companionship? Or what do you least want it to do? & 
    \textbf{Clarifying Boundaries}: Emphasizing ``assistance'' not ``replacement'' and understanding objections. \\
    \addlinespace
    & \textbf{Q33.} If we designed such an intelligent companion for you, what personality or traits would you most want it to possess? & 
    \textbf{Gaining Design Insights}: Making technological care more humanized by involving the user. \\
    \cmidrule(lr){1-3}

    % Module 6
    \textbf{(6) Information and Feedback} & 
    \textbf{Q34.} Regarding your care or daily life, are there specific, practical things we can help you improve? & 
    \textbf{Obtaining Practical Needs}: Pivot to concrete, actionable help for immediate issues. \\
    \addlinespace
    & \textbf{Q35.} What would you like to say to those who care for you (family, caregivers)? Gratitude, or unspoken requests? & 
    \textbf{Conveying Messages}: Providing a channel to express emotions to caregivers. \\
    \addlinespace
    & \textbf{Q36.} For those ``outside world'' people who are healthy and busy, do you have anything to tell them? About what is truly important. & 
    \textbf{Collecting Life Wisdom}: Inviting them to share universal life advice as mentors. \\
    \addlinespace
    & \textbf{Q37.} Based on your experience, do you have any advice for me, or others who wish to help people in similar situations? & 
    \textbf{Reverse Empowerment}: Letting the interviewee become a guide to enhance their sense of value. \\
    \addlinespace
    & \textbf{Q38.} Finally, what does today's conversation mean to you? Is there any part that you felt lighter after saying out loud? & 
    \textbf{Closing the Meaning Loop}: Reflecting on the value of the conversation itself. \\
    \cmidrule(lr){1-3}

    % Acknowledgement
    \textbf{Acknowledge--ment} & 
    ``I thank you again from the bottom of my heart. Thank you for your courage, wisdom, and trust in sharing such precious parts of your life with me. Your stories have deeply touched me, and I will carry them forward. This conversation is an invaluable gift. Please take care.'' & 
    \textbf{Sincere Thanks and Farewell}: Using sincere language to bring this profound connection to a complete close. \\
    
\end{longtable}

\newpage

\section{Qualitative Interview Results -- Statistical Summary of Participant Needs}

\begin{longtable}{@{} p{0.12\linewidth} p{0.30\linewidth} p{0.53\linewidth} @{}}
    \caption{Summary of Core Needs and Manifestations by Dimension} \label{tab:qualitative_results} \\
    \toprule
    \textbf{Dimension} & \textbf{Core Need} & \textbf{Specific Manifestations and Requests} \\
    \midrule
    \endfirsthead

    % Header for subsequent pages
    \multicolumn{3}{c}%
    {{\bfseries Table \thetable\ continued from previous page}} \\
    \toprule
    \textbf{Dimension} & \textbf{Core Need} & \textbf{Specific Manifestations and Requests} \\
    \midrule
    \endhead

    % Footer for pages except the last
    \bottomrule
    \multicolumn{3}{r}{{Continued on next page}} \\
    \endfoot

    % Footer for the last page
    \bottomrule
    \endlastfoot

    % --- Psychological and Emotional Dimension ---
    \textbf{Psychological \& Emotional Dimension} 
    & \textbf{1. Countering Existential Nihilism: Craving Deep ``Being-with''} \par
    \textit{Fear lies not merely in loneliness, but in the sense of void where one's existence is erased. The need is not for physical company, but for a ``Witness'' of the soul who can penetrate the barrier of isolation.} 
    & \begin{itemize}[leftmargin=*, nosep, after=\vspace{\baselineskip}]
        \item \textbf{Thirst for ``Authentic'' Dialogue}: Weary of pleasantries; suddenly posing essential questions like ``Do you think my life was worth living?'', seeking philosophical dialogue.
        \item \textbf{Extreme Longing for Touch}: Handshakes, hugs, stroking hair. This is not just comfort, but sensory evidence confirming one's physical existence through another's warmth.
        \item \textbf{Creating Micro-Rituals}: Fixed daily chat times or completing small tasks together to establish order and expectation in chaotic time, proving one is ``still present.''
    \end{itemize} \\
    \cmidrule{2-3}
    
    % Empty first column for continuation
    & \textbf{2. Seeking Ultimate Confirmation of Meaning: Closing the ``Life Narrative''} \par
    \textit{Need to weave scattered life fragments into a complete story with a beginning and an end, ensuring a ``fitting conclusion.''} 
    & \begin{itemize}[leftmargin=*, nosep, after=\vspace{\baselineskip}]
        \item \textbf{Becoming a ``Storyteller''}: Repeatedly narrating specific life fragments (often highlights or traumas), re-editing and imbuing meaning with each retelling to find the through-line of life.
        \item \textbf{Diversified Definition of ``Legacy''}: Focusing not on material wealth but on \textit{Spiritual Legacy}---a viewpoint, a chaotic act of kindness, or a character trait---and how it is inherited.
        \item \textbf{Seeking Recognition from Others}: Urgent desire to hear from significant others: ``Your life was remarkable,'' or ``You are important to me,'' serving as a stamp of approval on their autobiography.
    \end{itemize} \\
    \cmidrule{2-3}

    & \textbf{3. Processing Unfinished Business: Craving ``Completion'' and ``Closure''} \par
    \textit{Unfinished business acts as an ``unclosed wound'' on the soul, causing suffering greater than physical pain.} 
    & \begin{itemize}[leftmargin=*, nosep, after=\vspace{\baselineskip}]
        \item \textbf{Desire to Act as ``Messenger''}: Wishing to deliver an apology, gratitude, or long-buried love to resolve karma.
        \item \textbf{Completing Symbolic Actions}: Tearing up an old letter (letting go), organizing an album (summarizing), or gifting a personal item (continuing connection). External rituals for internal shifts.
        \item \textbf{Seeking ``Permission'' to Let Go}: Unable to leave due to responsibility; needing to hear loved ones say ``It's okay now'' to gain the ``moral permission'' to depart peacefully.
    \end{itemize} \\
    \cmidrule{2-3}

    & \textbf{4. Expressing Fundamental Fear \& Gaining ``Unconditional Holding''} \par
    \textit{Fear stems from the unknown, loss of control, and total separation. The comfort needed is not ``don't be afraid,'' but ``I am here, facing it with you.''} 
    & \begin{itemize}[leftmargin=*, nosep, after=\vspace{\baselineskip}]
        \item \textbf{Inquiry into the Process of Dying}: More than the afterlife, they fear the process of dignity loss---pain, incontinence, confusion.
        \item \textbf{Fear of Darkness and Silence}: Need for a nightlight or background music, not just for comfort, but to combat the absolute void symbolizing extinction.
        \item \textbf{Need for ``Unavoiding Sincerity''}: Forced optimism or avoidance of death by others deepens loneliness. They need courage from others to face fear and promise ``non-abandonment.''
    \end{itemize} \\
    \cmidrule{2-3}
    
    & \textbf{5. Maintaining Dignity and Self-Identity} \par
    \textit{Defending the final fortress of ``Who I Am'' when physical functions collapse.} 
    & \begin{itemize}[leftmargin=*, nosep, after=\vspace{\baselineskip}]
        \item \textbf{Insistence on Personal History}: Correcting misstatements about their story; insisting on specific titles.
        \item \textbf{Adherence to Aesthetics and Habits}: E.g., keeping hair neat. Micro-insistences are the final expression of ``Self-Style.''
        \item \textbf{Defining Privacy Boundaries}: Wishing to retain private space or information even in fragility, as a final stance on autonomy.
    \end{itemize} \\
    \midrule

    % --- Practical and Control Dimension ---
    \textbf{Practical \& Control Dimension} 
    & \textbf{1. Maintaining Autonomy \& Control: Establishing ``Minimal Order''} \par
    \textit{When macro-control over life is lost, control over micro-affairs becomes the sole fulcrum of existence.} 
    & \begin{itemize}[leftmargin=*, nosep, after=\vspace{\baselineskip}]
        \item \textbf{Extreme Granularity of Choice}: E.g., ``Should the cup be on the left or right?'', ``Song A or Song B?''.
        \item \textbf{Exercising Power through Refusal}: Refusing food, medicine, or visits. This may be the only domain where they can say ``No'' and compel obedience---a tragic display of subjectivity.
        \item \textbf{Creating Personal ``Rules''}: E.g., dictating the order of items, creating a small ``kingdom'' still under their rule.
    \end{itemize} \\
    \cmidrule{2-3}

    & \textbf{2. Arranging Afterlife Affairs: Extension of ``Post-mortem Will''} \par
    \textit{Not just for order, but to extend ``influence'' into the world after death---an attempt to transcend mortality.} 
    & \begin{itemize}[leftmargin=*, nosep, after=\vspace{\baselineskip}]
        % FIXED: Escaped the & in "Precision & Immutability"
        \item \textbf{Pursuit of Precision \& Immutability}: Exacting details for wills/funerals, as the final ``direction'' of their life script.
        \item \textbf{Distributing Symbolic Items}: Leaving a specific book or jewelry to a specific person with a reason. A final ``relational positioning'' and emotional expression.
        \item \textbf{Handling Digital Legacy}: Passwords, electronic albums. Fear that their ``soul'' in the digital world will become an untended ghost.
    \end{itemize} \\
    \midrule

    % --- Social and Existential Dimension ---
    \textbf{Social \& Existential Dimension} 
    & \textbf{1. Longing to be ``Witnessed'' \& Remembered: Fighting ``Ultimate Oblivion''} \par
    \textit{Knowing the body will perish, existence is pinned on others' memory/narration---pursuing ``Social Immortality.''} 
    & \begin{itemize}[leftmargin=*, nosep, after=\vspace{\baselineskip}]
        \item \textbf{Binding Name to Story}: Wishing to be remembered not as a name symbol, but for the story behind it.
        \item \textbf{Leaving Material Traces}: Handprints, recordings, writings---physical evidence of ``Having Been Here.''
        \item \textbf{Seeking a ``Keeper of Memory''}: Solemnly entrusting someone: ``Please remember this for me.''
    \end{itemize} \\
    \cmidrule{2-3}

    & \textbf{2. Establishing Final Symbolic Connections: A ``Poetic Farewell'' to the World} \par
    \textit{Condensed, metaphorical actions representing a final, affectionate dialogue with the world.} 
    & \begin{itemize}[leftmargin=*, nosep, after=\vspace{\baselineskip}]
        \item \textbf{Reunion with Natural Elements}: Seeing the last sunset, touching soil, hearing rain. Returning to life's origin to bid the universe farewell.
        \item \textbf{Transcendence through Art}: Listening to a life-summarizing song or poem. Art provides a sublime framework transcending personal pain.
        \item \textbf{Completing a ``Loop''}: E.g., returning to a childhood home, drawing a formal period on life.
    \end{itemize} \\
    \cmidrule{2-3}

    & \textbf{3. Integration and Reconciliation: Arriving at ``Final Unity''} \par
    \textit{Integrating light and shadow, love and hate, success and failure into an acceptable, complete life work.} 
    & \begin{itemize}[leftmargin=*, nosep, after=\vspace{\baselineskip}]
        \item \textbf{Self-Forgiveness}: Letting go of self-blame; admitting ``I did my best''; accepting the imperfect self.
        \item \textbf{Reconstructing Trauma Narratives}: Reinterpreting suffering as ``what made me strong'' or ``unique,'' finding positive meaning in pain.
        \item \textbf{Reconciling with Fate}: Shifting from the anger of ``Why me?'' to the peaceful acceptance of ``This is my unique life.''
    \end{itemize} \\
    \midrule

    % --- Spiritual and Transcendent Dimension ---
    \textbf{Spiritual \& Transcendent Dimension} 
    & \textbf{1. Exploring Ultimate Meaning: Questioning ``Existence'' Itself} \par
    \textit{Transcending personal narrative to consider the universal fate of the species, touching the core of philosophy/religion.} 
    & \begin{itemize}[leftmargin=*, nosep, after=\vspace{\baselineskip}]
        \item \textbf{Questioning \& Reconstructing Faith}: Believers may question God; atheists may ponder spirituality. Seeking the most personal, authentic ultimate belief.
        \item \textbf{Contemplating Cosmic Vastness}: Discussing stars/infinity. In a grand frame of reference, personal death becomes small, offering strange comfort.
        \item \textbf{Exploring the ``I Am'' behind ``Who I Am''}: Stripping away social roles/body to explore the essential core of consciousness.
    \end{itemize} \\
    \cmidrule{2-3}

    & \textbf{2. Seeking Trans-individual Connection: Merging into the ``Greater Whole''} \par
    \textit{Breaking the cage of individuation to feel part of the river of life, human civilization, or cosmic energy.} 
    & \begin{itemize}[leftmargin=*, nosep, after=\vspace{\baselineskip}]
        \item \textbf{Caring for Humanity's Future}: Extending care to community/earth/offspring. Transcending ``personal end'' by caring for the ``future.''
        \item \textbf{Hoping to be ``Nutrient''}: Hoping experiences serve as lessons for others---transforming from ``individual'' to ``nourishment.''
        \item \textbf{Experiencing ``Oneness''}: In meditation/dreams, feeling the dissolution of boundaries and merging with all things.
    \end{itemize} \\
    \cmidrule{2-3}

    & \textbf{3. Experiencing Peace and Letting Go} \par
    \textit{Not passive surrender, but an active, trust-filled handover---the final active achievement of life.} 
    & \begin{itemize}[leftmargin=*, nosep, after=\vspace{\baselineskip}]
        \item \textbf{From Language to Silence}: A ``Companionable Silence,'' feeling a peace beyond words.
        \item \textbf{Inward Withdrawal of Focus}: Turning from the external world to internal feelings, listening to the call from the depths of life.
    \end{itemize} \\
    
\end{longtable}

\section{IUI Design Principles for Palliative Care}
\begin{longtable}{@{} p{0.12\linewidth} p{0.30\linewidth} p{0.53\linewidth} @{}}
    \caption{IUI Design Principles: Pillars, Rationales, and Guidelines} \label{tab:design_principles} \\
    \toprule
    \textbf{Principle} & \textbf{Rationale (Source of Need)} & \textbf{Key Practice Guidelines} \\
    \midrule
    \endfirsthead

    % Header for subsequent pages
    \multicolumn{3}{c}%
    {{\bfseries Table \thetable\ continued from previous page}} \\
    \toprule
    \textbf{Principle} & \textbf{Rationale (Source of Need)} & \textbf{Key Practice Guidelines} \\
    \midrule
    \endhead

    % Footer for pages except the last
    \bottomrule
    \multicolumn{3}{r}{{Continued on next page}} \\
    \endfoot

    % Footer for the last page
    \bottomrule
    \endlastfoot

    % ==========================================
    % Pillar I
    % ==========================================
    \multicolumn{3}{l}{\textbf{Pillar I: Interaction Experience (Humanized Empathic Connection)}} \\
    \midrule

    \textbf{1. Proactive Empathy \& Validation} & 
    \raggedright \textit{Source: Psychological (Fear \& Holding)} \par
    Users need understanding, not preaching. Empty assurances deepen loneliness; only sincere emotional resonance builds trust. & 
    \begin{itemize}[leftmargin=*, nosep, after=\vspace{\baselineskip}]
        \item Analyze voice tone and behavioral patterns to identify emotions.
        \item Provide quiet companionship during sorrow; guide reminiscence during energetic moments.
        \item Prioritize emotional validation (e.g., ``That sounds very difficult for you'').
    \end{itemize} \\
    \cmidrule(lr){1-3}

    \textbf{2. User Sovereignty \& Empowerment} & 
    \raggedright \textit{Source: Practical \& Control (Autonomy)} \par
    In a body-out-of-control context, absolute control over interaction is the final stronghold for dignity and self-efficacy. & 
    \begin{itemize}[leftmargin=*, nosep, after=\vspace{\baselineskip}]
        \item Always seek explicit permission (e.g., ``Would you like to chat now?'').
        \item Allow users to easily pause or exit any interaction.
        \item Offer choices rather than issuing commands to enhance the sense of control.
    \end{itemize} \\
    \cmidrule(lr){1-3}

    \textbf{3. Seamless Invisibility} & 
    \raggedright \textit{Source: Psychological (Dignity \& Identity)} \par
    Users crave communication with an ``understanding presence,'' not a cold machine. Visible tech reminds them of their frailty and dependence. & 
    \begin{itemize}[leftmargin=*, nosep, after=\vspace{\baselineskip}]
        \item Adhere to a minimalist, calming interface design.
        \item Support natural language conversation.
        \item Ensure intuitive flows so users focus on the exchange, not the operation.
    \end{itemize} \\
    \cmidrule(lr){1-3}

    \textbf{4. Hyper-Usability \& Inclusion} & 
    \raggedright \textit{Source: Physical (Extreme Frailty)} \par
    Any operational difficulty translates into frustration and powerlessness, stripping confidence in using the system. & 
    \begin{itemize}[leftmargin=*, nosep, after=\vspace{\baselineskip}]
        \item Minimize interaction latency.
        \item Support multi-modal inputs (voice, light touch, gesture).
        \item Ensure the interface is clearly readable and easy to operate in any state.
    \end{itemize} \\
    \midrule

    % ==========================================
    % Pillar II
    % ==========================================
    \multicolumn{3}{l}{\textbf{Pillar II: System Function (Proactive Life Guardian)}} \\
    \midrule

    \textbf{5. Proactive Care \& Initiation} & 
    \raggedright \textit{Source: Psychological (Countering Nihilism)} \par
    Depression drains the energy to initiate. System proactivity is a pebble in the void, proving the user is still ``seen'' and cared for. & 
    \begin{itemize}[leftmargin=*, nosep, after=\vspace{\baselineskip}]
        \item Gently initiate care based on behavior (e.g., prolonged silence), time, or bio-data.
        \item Example: ``You seem quieter than usual today. Would you like to talk?''
    \end{itemize} \\
    \cmidrule(lr){1-3}

    \textbf{6. Life Narrative \& Witnessing} & 
    \raggedright \textit{Source: Social/Existential (Social Immortality)} \par
    The core demand to oppose ``ultimate oblivion.'' The system acts as the final listener and recorder, preserving their legacy. & 
    \begin{itemize}[leftmargin=*, nosep, after=\vspace{\baselineskip}]
        \item Use guided life-review dialogues to help users tell their stories.
        \item Automatically compile audio stories into digital legacies (biographies).
        \item Ensure materials are preserved and transferred according to their will.
    \end{itemize} \\
    \cmidrule(lr){1-3}

    \textbf{7. Meaning Construction} & 
    \raggedright \textit{Source: Psychological (Ultimate Confirmation)} \par
    Users need to integrate scattered life fragments into a meaningful whole to achieve a sense of closure and value. & 
    \begin{itemize}[leftmargin=*, nosep, after=\vspace{\baselineskip}]
        \item Guide positive life reviews through dialogue design.
        \item Help identify core values and spiritual legacies.
        \item Reinforce the sense of completion: ``My life has been meaningful.''
    \end{itemize} \\
    \cmidrule(lr){1-3}

    \textbf{8. Comfort \& Symptom Support} & 
    \raggedright \textit{Source: Physio-Psychological (Fear of Pain)} \par
    Physical pain exacerbates psychological fear and loss of dignity. Managing comfort is the foundation for maintaining composure. & 
    \begin{itemize}[leftmargin=*, nosep, after=\vspace{\baselineskip}]
        \item Seamlessly control smart home devices (lighting, thermostat).
        \item Play distraction content (music, meditation) to alleviate distress.
        \item Provide simple symptom reporting tools for the care team.
    \end{itemize} \\
    \cmidrule(lr){1-3}

    \textbf{9. Social Connection Bridge} & 
    \raggedright \textit{Source: Social (Symbolic Connection)} \par
    Users desire a final meaningful interaction and farewell with the world, rather than departing in isolated silence. & 
    \begin{itemize}[leftmargin=*, nosep, after=\vspace{\baselineskip}]
        \item Simplify the process of contacting family and friends.
        \item Intelligently connect to community volunteers or support groups.
        \item Ensure all external links are built on established trust.
    \end{itemize} \\
    \midrule

    % ==========================================
    % Pillar III
    % ==========================================
    \multicolumn{3}{l}{\textbf{Pillar III: Tech Ethics (Sacred Responsibility)}} \\
    \midrule

    \textbf{10. Absolute Reliability} & 
    \raggedright \textit{Source: Emotional Vulnerability} \par
    When the system is the sole sustenance, a failure is like being abandoned by a friend. Such betrayal is devastating. & 
    \begin{itemize}[leftmargin=*, nosep, after=\vspace{\baselineskip}]
        \item System must possess extreme robustness and reliability.
        \item Zero tolerance for unresponsiveness or glitches.
        \item Pursue perfection from system architecture to maintenance.
    \end{itemize} \\
    \cmidrule(lr){1-3}

    \textbf{11. Sacred Data Custody} & 
    \raggedright \textit{Source: Social/Existential (Legacy)} \par
    Digital legacy is the extension of life. Failing this trust is equivalent to denying the value of their existence and final will. & 
    \begin{itemize}[leftmargin=*, nosep, after=\vspace{\baselineskip}]
        \item Employ highest-level encryption and security.
        \item Faithfully and reliably execute digital legacy transfer instructions post-mortem.
        \item Treat this as the ultimate respect for the user's life.
    \end{itemize} \\
    \cmidrule(lr){1-3}

    \textbf{12. Privacy \& Ethical Supremacy} & 
    \raggedright \textit{Source: Psychological (Dignity)} \par
    At the end of life, users expose their most fragile thoughts. Protecting privacy preserves their final personality boundary. & 
    \begin{itemize}[leftmargin=*, nosep, after=\vspace{\baselineskip}]
        \item All sensitive data processing must be transparent and ethical.
        \item Strictly define data boundaries to prevent abuse.
        \item Especially protect intimate confessions and after-death arrangements.
    \end{itemize} \\

\end{longtable}

\section{The Roles and Applications of IUI in Palliative Care}

% 5 Columns: ID, Name, Description & Value, Precautions, Mapped Core Need
% Adjusted widths to fit page: 5% + 12% + 38% + 20% + 20% = 95%
\begin{longtable}{@{} p{0.05\linewidth} p{0.12\linewidth} p{0.38\linewidth} p{0.20\linewidth} p{0.20\linewidth} @{}}
    \caption{IUI Functions, Values, and Core Need Mapping} \label{tab:iui_roles} \\
    \toprule
    \textbf{No.} & \textbf{Function} & \textbf{Description \& Value} & \textbf{Precautions} & \textbf{Mapped Core Need} \\
    \midrule
    \endfirsthead

    % Header for subsequent pages
    \multicolumn{5}{c}%
    {{\bfseries Table \thetable\ continued from previous page}} \\
    \toprule
    \textbf{No.} & \textbf{Function} & \textbf{Description \& Value} & \textbf{Precautions} & \textbf{Mapped Core Need} \\
    \midrule
    \endhead

    % Footer for pages except the last
    \bottomrule
    \multicolumn{5}{r}{{Continued on next page}} \\
    \endfoot

    % Footer for the last page
    \bottomrule
    \endlastfoot

    % IUI 1
    IUI1 & Deep Dialogue Partner & 
    \textbf{Description:} Based on LLMs, engages in preset-free, empathic dialogue covering memories, feelings, and life reflections. \newline
    \textbf{Value:} Provides a 24/7 non-judgmental listener to counter loneliness and confirm the user's existence through active interaction. & 
    Must clearly state this is AI interaction, not a replacement for professional psychological counseling. & 
    \textbf{Psych-Emo.1:} Countering Existential Nihilism (Soul Witness) \newline
    \textbf{Psych-Emo.4:} Unconditional Holding (Non-judgmental Company) \\
    \cmidrule(lr){1-5}

    % IUI 2
    IUI2 & Remote Embrace Experience & 
    \textbf{Description:} Integrates with wearable devices to simulate warmth and heartbeat-like pressure when triggered by a remote loved one. \newline
    \textbf{Value:} Provides symbolic tactile comfort when physical contact is impossible, reinforcing the sense of love and connection. & 
    Clarify this is simulated touch; cannot replace real hugs. Requires user willingness to wear the device. & 
    \textbf{Psych-Emo.1:} Countering Existential Nihilism (Symbolic Being-with) \newline
    \textbf{Social-Exist.2:} Symbolic Connection (Deep Affectionate Farewell) \\
    \cmidrule(lr){1-5}

    % IUI 3
    IUI3 & Life Memoir Creation & 
    \textbf{Description:} Guides users to tell life stories and auto-compiles them with photos into a structured ebook or audio biography. \newline
    \textbf{Value:} Helps users organize life trajectory, visualize core values, complete the ``Life Narrative'' loop, and create an eternal memorial. & 
    Reviewing may trigger strong emotions. User retains full control over content and pace. & 
    \textbf{Psych-Emo.2:} Ultimate Confirmation of Meaning (Narrative Closure) \newline
    \textbf{Social-Exist.1:} Witnessing \& Remembering (Tangible Anti-oblivion) \\
    \cmidrule(lr){1-5}

    % IUI 4
    IUI4 & Wish \& Regret Assistant & 
    \textbf{Description:} Assists in finding information or drafting messages when users mention unfinished wishes (e.g., contacting someone). \newline
    \textbf{Value:} Transforms mental burdens into actionable steps, facilitating peace and reconciliation through low-threshold assistance. & 
    Final execution right lies solely with the user. System only assists and ensures privacy. & 
    \textbf{Psych-Emo.3:} Unfinished Business (Actionable Healing) \newline
    \textbf{Social-Exist.3:} Integration \& Reconciliation (Inner Unity) \\
    \cmidrule(lr){1-5}

    % IUI 5
    IUI5 & Emotional Soothing Menu & 
    \textbf{Description:} Proactively offers soothing options (chat, meditation, scenery) upon detecting unease. \newline
    \textbf{Value:} Provides a controllable ``comfort toolbox'' to help stabilize emotions and rebuild a sense of safety during anxiety. & 
    Not a substitute for medical care. Must guide to emergency call if severe physical distress occurs. & 
    \textbf{Psych-Emo.4:} Unconditional Holding (Present Comfort) \newline
    \textbf{Psych-Emo.1:} Countering Nihilism (Support in Fear) \\
    \cmidrule(lr){1-5}

    % IUI 6
    IUI6 & Personalized Env. Steward & 
    \textbf{Description:} Learns habits to auto-coordinate environment (light, music, temp) and remind caregivers. \newline
    \textbf{Value:} Maintains personal habits and order through attentive care, preserving dignity and control amidst physical decline. & 
    Control of critical medical equipment requires extra authorization from user/staff. & 
    \textbf{Psych-Emo.5:} Dignity \& Identity (Preserving ``Who I Am'') \newline
    \textbf{Practical-Control.1:} Autonomy (Minimal Order) \\
    \cmidrule(lr){1-5}

    % IUI 7
    IUI7 & Digital Legacy Guide & 
    \textbf{Description:} Guides users to record digital assets (photos, accounts) and assign inheritance wishes securely. \newline
    \textbf{Value:} Allows users to dispose of digital property according to their will, ensuring memory preservation and transfer. & 
    All info encrypted. User advised to inform a trusted person of this arrangement. & 
    \textbf{Practical-Control.2:} Post-mortem Affairs (Will Extension) \newline
    \textbf{Social-Exist.1:} Witnessing \& Remembering (Digital Footprint) \\
    \cmidrule(lr){1-5}

    % IUI 8
    IUI8 & Virtual Poetic Journey & 
    \textbf{Description:} Uses Screen/VR to immerse users in inaccessible places (e.g., childhood home, nature). \newline
    \textbf{Value:} Facilitates a ``Poetic Farewell'' to the world, fulfilling unmet wishes spiritually for solace and transcendence. & 
    May cause dizziness. Provide immediate stop option and alternatives. & 
    \textbf{Social-Exist.2:} Symbolic Connection (Poetic Farewell) \newline
    \textbf{Spiritual.3:} Peace \& Letting Go (Wish Fulfillment) \\
    \cmidrule(lr){1-5}

    % IUI 9
    IUI9 & Life Meaning Exploration & 
    \textbf{Description:} Provides diverse cross-cultural, philosophical, and religious perspectives when users ponder life/death. \newline
    \textbf{Value:} Accompanies deep thinking, helping users find their own peace and answers through diverse wisdom. & 
    Clarify these are intellectual resources, not absolute truths or religious preaching. & 
    \textbf{Spiritual.1:} Ultimate Meaning (Existential Inquiry) \newline
    \textbf{Psych-Emo.2:} Meaning Confirmation (Philosophical Material) \\
    \cmidrule(lr){1-5}

    % IUI 10
    IUI10 & Serenity Moment Guide & 
    \textbf{Description:} Plays personalized guided meditation or nature sounds based on preferences. \newline
    \textbf{Value:} Creates a quiet soundscape during inner turmoil or rest, supporting focus on the present and preparation for ``letting go.'' & 
    If user falls asleep, system should auto-fade volume and enter silent guard mode. & 
    \textbf{Spiritual.3:} Peace \& Letting Go (Creating Serenity) \newline
    \textbf{Psych-Emo.4:} Unconditional Holding (Soothing Anxiety) \\

\end{longtable}

\end{document}